\documentclass[pdflatex,tex,twocolumn,epjc3]{svjour3}
\RequirePackage[T1]{fontenc}
\usepackage{color}
\newcommand{\beqa}{\begin{eqnarray}}
\newcommand{\eeqa}{\end{eqnarray}}
\usepackage{float}
\usepackage{epsfig}
\usepackage{graphicx}
\usepackage{latexsym}

\usepackage{mathrsfs}
\usepackage{dcolumn}% Align table columns on decimal point
\usepackage{bm}%

\usepackage{indentfirst}
\usepackage{amssymb,dcolumn}
\usepackage{latexsym}

\usepackage{setspace}
\usepackage{graphicx,float}
\usepackage[colorlinks,linkcolor=blue,citecolor=blue,urlcolor=blue,hyperindex]{hyperref}
\usepackage{color}
\usepackage{slashed}
\usepackage{amsmath}

\journalname{Eur. Phys. J. C}
\begin{document}

\title{Constraining deviations from $\Lambda$CDM in the Hubble expansion rate}

\author{Yupeng Yang\thanksref{addr1,e1}}
\thankstext{e1}{e-mail: ypyang@qfnu.edu.cn}

\institute{School of Physics and Physical Engineering, Qufu Normal University, Qufu, Shandong, 273165, China \label{addr1}}

\date{Received: date / Accepted: date}

\maketitle

\begin{abstract}

The $\Lambda$CDM cosmolgical model has long been regarded as highly successful in accurately describing 
a wide range of astronomical observations. 
However, numerous observational findings have also provided hints of discrepancies from the predictions of the $\Lambda$CDM framework. 
We explore a phenomenological model that quantifies the deviation of the Hubble expansion rate from the standard scenario, 
which is expressed as $H^{2}(z) = H^{2}_{\rm \Lambda CDM}(\Omega_m, z)[1+\delta(z)]$. 
We consider three distinct forms for the deviation parameter $\delta(z)$: in model I, $\delta(z)=\delta_c$; in model II, $\delta(z)=\delta_{c}z/(1+z)$, and in model III, 
$\delta(z)=\delta_{c}{\rm ln}(1+z)$. Here, $\delta_c$ represents a constant value. 
We utilize a comprehensive set of observational data to constrain the models. 
Our results show that for most combined datasets, $\delta_c$ tends to take on negative values for models I and II, 
while consistently taking positive values in model III. Furthermore, we find that both 
models I and II remain consistent with the standard $\Lambda$CDM model across all datasets examined. 
In contrast, model III exhibits a significant deviation from the $\Lambda$CDM model, exceeding $2\sigma$ for the full 
combined datastes. 
The AIC indicates that models I and II are consistent with the $\Lambda$CDM model, whereas model III is 
preferred over the standard $\Lambda$CDM model, with the $\Lambda$CDM model being disfavored for the combined datasets 
DESI BAO + CMB + CC + DESY5. These results suggest that the Hubble expansion rate likely deviates from the standard 
$\Lambda$CDM prediction.

\end{abstract}

\maketitle

\section{Introduction} 

In the widely accepted standard cosmological model, known as $\Lambda$CDM, the Universe is predominantly composed of 
cold dark matter (CDM) and a cosmological constant $\Lambda$ as dark energy, 
with the latter characterized by a constant density and a fixed equation of state 
$w_{\Lambda}=p_{\Lambda}/\rho_{\Lambda}=-1$. This model has proven to be highly successful, 
providing an excellent fit to a wide range of observational data~\cite{2020A&A...641A...6P,Wallisch2019}. 
Despite its significant achievements, however, numerous observational findings have indicated discrepancies with the $\Lambda$CDM model~\cite{Perivolaropoulos:2021jda,Bull:2015stt,Hu:2023jqc,Li:2025ops,Vagnozzi:2023nrq,Vagnozzi:2019ezj}. 
Recently, the Dark Energy Spectroscopic Instrument (DESI) released its first and third-year data (DR1 and DR2)~\cite{DESI:2024mwx,DESI:2025zgx}. When combined with data from 
the Cosmic Microwave Background (CMB) and Type Ia supernovae (SNIa), these datasets suggest a preference for a dynamical dark energy model 
with $w=w(z)$, thereby revealing a deviation from the standard $\Lambda$CDM framework~\cite{DESI:2025zgx}~\footnote{It should be noted, however, 
that the evidence for dynamical dark energy remains debated~\cite{Colgain:2025nzf,Huang:2025som,Notari:2024zmi,Gialamas:2024lyw,Efstathiou:2025tie}, and the signal inferred from 
these combined datasets may be biased by low-redshift supernova observations.}. 

The Hubble tension problem has emerged as a major challenge to the $\Lambda$CDM model, arising from a significant 
discrepancy in measurements of the Hubble constant ($H_0$). Local observations, such as those from Type Ia supernovae and 
Cepheid variable stars, which do not dependent the cosmological model, 
yield $H_0 = 73.04 \pm 1.04~\rm{km~s^{-1}~Mpc^{-1}}$~\cite{2009ApJ...699..539R,Breuval:2024lsv}. 
In contrast, early-universe constraints based on 
the $\Lambda$CDM model, which are derived from cosmic microwave background (CMB) and baryon acoustic oscillations (BAO) data, 
give a value of $67.4 \pm 0.5~\rm{km~ s^{-1} ~ Mpc^{-1}}$~\cite{2020A&A...641A...6P}. This disagreement reaches a statistical significance of $\sim 5\sigma$, 
posing a serious challenge to the standard cosmological paradigm. Numerous theoretical approaches have been proposed to resolve this tension, 
see, e.g., Refs.~\cite{Yang:2025vnm,Poulin:2018cxd,Hu:2023jqc,Brito:2024bhh,DiValentino:2021izs,Elizalde:2021kmo,Khurshudyan:2024gpn,CosmoVerseNetwork:2025alb,Zhang:2025dwu}.

Numerous alternative cosmological models, distinct from the $\Lambda$CDM framework, have been put forward to better 
align with observational data, and many of these models demonstrate a notable preference for the available data 
when compared to the $\Lambda$CDM model
~\cite{8ync-vrtz,SolaPeracaula:2017esw,Pan:2025qwy,SolaPeracaula:2016qlq,Sola:2016jky,Li:2025dwz,Du:2025xes,Feng:2025wbz,Guedezounme:2025wav,Millano:2025vjo,Capozziello:2025kws,Kumar:2025etf,Wu:2025wyk,Wu:2024faw,Wang:2025znm,Wang:2023ros,Li:2024qso,Li:2025owk,Adhikary:2025khr,Nojiri:2025low,Odintsov:2025jfq,Odintsov:2024woi,Braglia:2025gdo,RoyChoudhury:2024wri,RoyChoudhury:2025dhe,RYSKIN2015258,Wu:2025vfs,Zhou:2025nkb,Wolf:2025acj,RoyChoudhury:2025iis,Alam:2025epg,Li:2025muv,Du:2025csv}. 
Regardless of the specific cosmological models, all those models that fit the relevant 
data better than the $\Lambda$CDM model represent deviations from the standard $\Lambda$CDM model. Of course, the extent and manner 
of deviation vary among different models. Here, based on the ideas in Refs.~\cite{Samsing_2012,Hojjati:2013oya}, 
we formally consider a relatively universal model. 
This model can directly describe the degree of deviation of Hubble expansion rate from the standard $\Lambda$CDM model. 
For a flat universe, the specific formulation of this model is given by $H^{2}(z)=H^{2}_{\rm \Lambda CDM}(\Omega_{m},z)[1+\delta(z)]$. 
Here, $H^{2}_{\rm \Lambda CDM}$ denotes the Hubble expansion rate as predicted by the standard $\Lambda$CDM model, 
while $\delta(z)$ encapsulates the deviation 
from the standard Hubble expansion rate. The authors in Refs.~\cite{Samsing_2012,Hojjati:2013oya} focused on 
constraining $\delta(z)$ using cosmic microwave background (CMB) data, which primarily probe the expansion rate of the early universe. 
They argued that any deviation of the Hubble expansion rate from the $\Lambda$CDM model would imply additional contributions 
from factors such as early dark energy or extra relativistic degrees of freedom. 
In light of recent advancements in various cosmological observations, we revisit the deviation of the Hubble expansion rate 
from the standard $\Lambda$CDM model in this study. 
In contrast to the approaches adopted in previous studies~\cite{Samsing_2012,Hojjati:2013oya}, our investigation explores 
three distinct parameterizations of $\delta(z)$. Our final results reveal substantial deviations ($\delta(z) \neq 0$) at a notably 
higher $3\sigma$ confidence level for certain models and combined datasets.

This paper is structured as follows. In Sec.~\ref{sec:basic}, we begin by providing an introduction to the models under investigation. 
Moving on to Sec.~\ref{sec2}, we present the datasets utilized in our analysis, while the final constraints on the pertinent parameters 
are presented in Sec.~\ref{cons}. A possible cosmological model is proposed in Sec.~\ref{model}, 
and finally, our conclusions are summarized in Sec.~\ref{con}.

%%%%%%%%%%%%%%%%%%%%%%%%%%%%%%%%%%%%%%%%%%%%%%%

\section{The specific models examined throughout this research}
\label{sec:basic}

For a flat universe, the Hubble expansion rate within the $\Lambda$CDM model can be expressed as

\beqa
H^{2}_{\rm \Lambda CDM}=H^{2}_{0}\left[\Omega_{m}(1+z)^{3}+\Omega_{\Lambda}+\Omega_{r}(1+z)^{4}\right]
\eeqa
where $\Omega_{m}$, $\Omega_{\Lambda}$, and $\Omega_{r}$ represents the fraction of the matter, dark energy (cosmological constant), 
and radiaton energy density at present, and they 
satisfy the condition $\Omega_{m}+\Omega_{\Lambda}+\Omega_{r}=1$. Following previous works and considering deviations from $\Lambda$CDM model, 
we consider the model as~\cite{Samsing_2012,Hojjati:2013oya}

\beqa
&&H^{2}=H^{2}_{\rm \Lambda CDM}[1+\delta(z)] \nonumber \\
&&=H^{2}_{0}\left[\Omega_{m}(1+z)^{3}+\Omega_{\Lambda}+\Omega_{r}(1+z)^{4}\right][1+\delta(z)].
\label{eq:basic}
\eeqa

In Eq.(\ref{eq:basic}), $\delta(z)$ quantifies the deviation of our considered model from the $\Lambda$CDM model. 
Generally, there is no fundamental law that restricts $\delta(z)$ from varying with time. Consequently, 
it can be a function of redshift, indicating that the deviation can manifest differently at various epochs of 
the Universe. The authors of~\cite{Samsing_2012,Hojjati:2013oya} have explored $\delta(z)$ in a model-independent manner, 
primarily focusing on constraints from CMB observations, which probe the early universe. In contrast, here we consider three specific models:

\[
\delta(z) = \left\{
\begin{aligned}
&\delta_{c}                  &&\text{Model I}    \\
&\delta_{c}\frac{z}{1+z}     &&\text{Model II}   \\
&\delta_{c}{\rm ln}(1+z)     &&\text{Model III}
\end{aligned}
\right.
\]
here, $\delta_c$ is treated as a constant parameter. Consequently, Model I describes a redshift-independent deviation. In contrast, 
models II and III incorporate redshift dependence and exhibit structural similarities to well-known dynamic dark energy parameterizations, 
specifically the Chevallier-Polarski-Linder (CPL)~\cite{cpl_1,cpl_2} and Logarithmic 
models~\cite{Tripathi:2016slv,Efstathiou:1999tm,Silva:2011pc,Feng_2011}, respectively. Importantly, the deviation parameter 
$\delta(z)$ can be equivalently interpreted as a non-cosmological constant form of dark energy~\cite{Samsing_2012,Hojjati:2013oya}. 
Notably, for all cases considered, the Hubble expansion rate predicted by our models reduces to the standard 
$\Lambda$CDM prediction when $\delta(z)=0$ (or equivalently, when $\delta_{c}=0$). 
Unlike previous studies~\cite{Samsing_2012,Hojjati:2013oya} in which only CMB data was used to constraint $\delta(z)$, 
we utilize multiple distinct datasets to constrain the models; these datasets are described in detail 
in the following section.

\section{The datasets}
\label{sec2}

%%%%%%%%%%%%%%%%%%%%%%   BAO  %%%%%%%%%%%%%%%%%%%%%
\subsection{Baryon acoustic oscillation}

The Dark Energy Spectroscopic Instrument (DESI) has released its three-year data (DR2), 
including six distinct classes of traces~\cite{DESI:2025zgx}:
the Bright Galaxy Sample (BGS,$z_{\rm eff}=0.295$), the Luminous Red Galaxy Sample (LRG1 and LRG2, $z_{\rm eff}=0.510$ and 0.706), 
the Emission Line Galaxy Sample (ELG2, $z_{\rm eff}=1.321$), 
the combined LRG and ELG Sample (LRG3+ELG1, $z_{\rm eff}=0.934$), 
the Quasar Sample (QSO, $z_{\rm eff}=1.484$) and the Lyman-$\alpha$ Forest Sample (Ly$\alpha$, $z_{\rm eff}=2.330$). 
The data released by the DESI collaboration are presented in the form of $D_{\rm M,H,V}/r_{d}$ 
and $D_{\rm M}/D_{\rm H}$. In the context of a homogeneous and isotropic cosmology, 
the transverse comoving distance, $D_{\rm M}(z)$, can be written as~\cite{2020A&A...641A...6P},

\beqa
D_{\rm M}(z) = \frac{c}{H_0}\int^{z}_{0}\frac{dz^{'}}{H(z^{'})/H_0}
\eeqa
where $c$ is the light speed and $H(z)$ is the Hubble parameter. The distance 
variable is defined as $D_{\rm H}(z)=c/H(z)$, and then the angle-averaged distance $D_{\rm V}$ can be written 
as,~\footnote{In Ref.~\cite{Li:2025htp}, the authors investigated deviations between the angular diameter distance and 
luminosity distance relations from their standard cosmological forms. In this work, however, such deviations are not considered.}

\beqa
D_{\rm V}(z) = \left[zD_{\rm M}(z)^{2}D_{\rm H}(z)\right]^{\frac{1}{3}}.
\eeqa

For the drag-epoch sound horizon $r_d$, we adopt the approximated form given in~\cite{DESI:2024mwx,DESI:2025zgx},

\beqa
r_{d} = \frac{147.05}{\rm Mpc}\left(\frac{\omega_{m}}{0.1432}\right)^{-0.23}\left(\frac{N_{\rm eff}}{3.04}\right)^{-0.1}
\left(\frac{\omega_{b}}{0.02236}\right)^{-0.13}. 
\eeqa
Here, $\omega_{m,b}=\Omega_{m,b}h^2$ with the reduced Hubble constant $h=H_{0}/(100~\rm km~s^{-1}~Mpc^{-1})$. 
We adopt the standard value of $N_{\rm eff} = 3.04$ for the effective number of relativistic degrees of freedom 
throughout our calculations.  

The $\chi^{2}$ statistic for DESI BAO data is given by

\beqa
\chi^{2}_{\rm BAO}=\sum_{i}\Delta D_{i}^{T}{\rm Cov}^{-1}_{\rm BAO} \Delta D_{i}
\eeqa
where $\Delta D_{i} = D_{i}^{th}-D_{i}^{obs}$, and the covariance matrix ${\rm Cov}_{\rm BAO}$ can be written as~\cite{Li:2024hrv}

\[{\rm Cov}_{\rm BAO}=
\begin{bmatrix}
\sigma^{2}_{1} & r\sigma_{1}\sigma_{2} \\
r\sigma_{1}\sigma_{2} & \sigma^{2}_{2} \\
\end{bmatrix}
\]
Here, $r$ is the correlation coefficient between the observation data, which has been provided in DR2~\cite{DESI:2025zgx}.

%%%%%%%%%%%%%%%%%%%%%%%%   CMB  %%%%%%%%%%%%%%%%%%%%%%%%%%%%%%%%%%%%
\subsection{Cosmic microwave background}

The Cosmic Microwave Background (CMB) data has been extensively employed to place constraints on a diverse range of 
cosmological models, with the angular power spectrum data being the most commonly utilized. However, 
beyond angular power spectrum data, distance priors can also provide constraints comparable to those from full power spectrum 
analysis when investigating models that modify $\Lambda$CDM at low redshifts. This is because, in such scenarios, 
the overall shape of the power spectrum remains largely unaffected, while only geometric changes influence CMB observations 
\cite{Chen:2018dbv,Wang:2007mza,Zhai:2019nad}. Consequently, CMB distance priors serve as effective tools for constraining 
cosmological models (see, e.g., Refs. \cite{Yang:2025vnm,Yang:2025boq,Zhai:2018vmm,Li:2024hrv,Jia:2025prq,Rezaei:2024vtg,Sohail:2024oki}). 
For the purposes of our study, 
rather than relying on the angular power spectrum data, we opt to utilize the distance prior parameters. 
These include the shift parameter $R$, the acoustic scale $l_{\rm A}$ 
and the baryon density $\Omega_{\rm b}h^{2}$, all derived from the Planck 2018 data, 
to constrain our parameters~\cite{2020A&A...641A...6P,Chen:2018dbv}. 
The shift parameter and acoustic scale can be espressed as~\cite{2020A&A...641A...6P,Liu:2018kjv,Xu:2016grp,Feng_2011},

\beqa
&&R=\frac{1+z_\star}{c}D_{\rm A}(z_{\star})\sqrt{\Omega_{m}H^{2}_0} \\ 
&&l_{\rm A}=(1+z_{\star})\frac{\pi D_{\rm A}(z_{\star})}{r_{s}(z_\star)}.
\eeqa
Here, angular diameter distance $D_A$ is defined as $D_{\rm A}=D_{\rm M}/(1+z)$. $z_\star$ is the redshift at the epoch of photon decoupling, 
and we adopt the approximate proposed by~\cite{Hu:1995en} 

\beqa
z_{\star} =1048[1+0.00124(\Omega_{b}h^{2})^{-0.738}][1+g_{1}(\Omega_{m}h^{2})^{g_{2}}]
\eeqa
where 

\beqa
g_{1}=\frac{0.0738(\Omega_{b}h^{2})^{-0.238}}{1+39.5(\Omega_{b}h^{2})^{0.763}},~ 
g_{2}=\frac{0.560}{1+21.1(\Omega_{b}h^{2})^{1.81}}
\eeqa

The quantity $r_s$ denotes the comoving sound horizon and can be written as~\cite{DESI:2024mwx},

\beqa
r_{s}(z)=\frac{c}{H_{0}}\int^{1/(1+z)}_{0}\frac{da}{a^{2}H(a)\sqrt{3(1+\frac{3\Omega_{b}h^{2}}{4\Omega_{\gamma}h^{2}}a)}}.
\eeqa

In this expression, $a=1/(1+z)$, and $(\Omega_{\gamma}h^{2})^{-1}=42000(T_{\rm CMB}/2.7{\rm K})^{-4}$, 
with $T_{\rm CMB}=2.7255\rm K$. 

The $\chi^{2}$ statistic for the CMB data is given by

\beqa
\chi^{2}_{\rm CMB}=\Delta X^{T}{\rm Cov}^{-1}_{\rm CMB} \Delta X
\eeqa
where $\Delta X=X-X^{\rm obs}$ is a vector with $X^{\rm obs}=(R,l_{\rm A},\Omega_{b}h^{2})$, 
and we utilized the values of $X^{\rm obs}$ and inverse covariance matrix ${\rm Cov}^{-1}_{\rm CMB}$ 
from Planck 2018 data given by, e.g., Ref.~\cite{Zhai:2018vmm}.
\footnote{Note that the covariance matrices in Refs.~\cite{Zhai:2018vmm} 
and \cite{Chen:2018dbv} differ slightly because they are derived from different data chains. 
We performed tests to assess this discrepancy and confirmed that the slight differences between their covariance matrices 
have a negligible impact on our final results.  
}
Although the covariance matrix is based on the flat $\Lambda$CDM model, 
its influence on our final results is less significant. Regarding this issue, one can refer to, 
e.g., Refs.~\cite{Wang:2007mza,Zhai:2019nad}, 
in which the authors investigated the influences of different dark energy models on the derived CMB distance priors 
and fond that for most models, the CMB distance priors are consistent. In this work, we consider 
models that have slight deviations from the $\Lambda$CDM model; therefore, the CMB distance priors 
derived from $\Lambda$CDM model are sufficient for our purposes. However, it should be noted that more accuracy constraints 
on our models from CMB data should stem from the use of full CMB data. 
%%%%%%%%%%%%%%%%%%%%%%%%   CMB  %%%%%%%%%%%%%%%%%%%%%%%%%%%%%%%%%%%%

%%%%%%%%%%%%%%%%%%%%%%%%%%% CC  %%%%%%%%%%%%%%%%%%%%%%%%%%%%%%%%%%%%
\subsection{Hubble rate from cosmic chronometers}

The Hubble rate can be determined through the cosmic chronometer (CC) method~\cite{Jimenez:2001gg,Moresco:2022phi,Jimenez:2023flo}. 
This technique primarily focuses on measuring the differential age evolution of the Universe, denoted as $dt$, 
within a specific redshift interval $dz$~\cite{Jimenez:2001gg}. In practical applications, 
this is achieved by analyzing a carefully selected sample of massive and passively evolving galaxies. 
The Hubble rate, as determined via the cosmic chronometer (CC) method, can be formulated as follows~\cite{Jimenez:2001gg}:

\beqa
H(z)=-\frac{1}{1+z}\frac{\Delta z}{\Delta t}.
\eeqa

The $\chi^{2}$ statistic for the CC data is given by

\beqa
\chi^{2}_{\rm CC}=\sum_{i}\frac{(H(z_i)-H_{\rm obs}(z_i))^{2}}{\sigma^{2}_i}
\eeqa
where we employ 31 CC data points, sourced from various 
Refs.~\cite{Li:2019nux,2010JCAP...02..008S,2012JCAP...08..006M,Moresco:2016mzx,Moresco:2015cya,Ratsimbazafy:2017vga,2014RAA....14.1221Z,Singirikonda:2020ieg}, 
to impose constraints on our model parameters. 
It should be noted that, in this analysis, we have neglected the potential correlations between the observational data points, 
which should has a minor influence on our final results.\footnote{For instance, results from Ref. \cite{Moresco:2024wmr} 
show that the difference in the mean values of parameters (e.g., $H_0$) between analyses that include and exclude 
correlations in CC data (accounting for statistical errors and various systematic errors) is less than 1\%.}  
In this work, we only consider statistical errors in our calculations. In fact, other factors can affect CC results 
\cite{Moresco:2020fbm,Moresco:2024wmr}; thus, for precise calculations, one should account for both statistical 
and various systematic errors. Nevertheless, the uncertainties in CC data are large compared to other datasets used in this work. 
As such, correlations in CC data or variations in systematic errors have negligible impacts on our final results, 
since the primary constraints come from other datasets. 
%%%%%%%%%%%%%%%%%%%%%%%%%%% CC  %%%%%%%%%%%%%%%%%%%%%%%%%%%%%%%%%%%%

%%%%%%%%%%%%%%%%%%%%%%% SN Ia  %%%%%%%%%%%%%%%%%%%%%%%%%%%%%%%%%%

\subsection{Type Ia supernova}

As a well-established standard candle, type Ia supernovae (SNIa) serve as a crucial tool 
for constraining cosmological models~\cite{SupernovaSearchTeam:1998fmf,SupernovaCosmologyProject:1998vns}. 
Among the key observables derived from SNIa data, the luminosity distance  $d_{L}(z)$ stands out as particularly significant. 
In a flat universe, the luminosity distance can be expressed mathematically as,~\footnote{In practice, the luminosity distance 
is calculated using the expression $d_{L}(z_{\rm hel}) = (1+z_{\rm hel})\int^{z_{\rm cmb}}_{0}\frac{dz^{'}}{H(z^{'})/H_0}$. 
Here, $z_{\rm hel}$ is the heliocentric redshift, and $z_{\rm cmb}$ refers to the CMB-corrected redshift. 
Both of values are provided in the released public datasets.}
 
\beqa
d_{L}(z) = (1+z)\int^{z}_{0}\frac{dz^{'}}{H(z^{'})/H_0}.
\eeqa

The distance modules $\mu(z)$ is then defined as,

\beqa
\mu(z) = 5{\rm log}_{10}d_{L}(z)+25.
\eeqa
 
For SNIa data, the observed apparent magnitude, $m_{\rm obs}$, is related the observed distance modules by

\beqa
m_{\rm obs} = \mu_{\rm obs} + M
\eeqa
where \(M\) represents the absolute magnitude (in B-band), which can be calibrated using alternative methods. 
In this work, however, we opt not to utilize a pre-calibrated value of $M$ as suggested in, for instance, 
Refs.~\cite{Camarena:2019moy,Camarena:2021jlr,Kang:2019azh,vonMarttens:2025dvv,Lemos:2025qyh,Riess:2019cxk,Chander:2025bml,Camarena:2023rsd}, 
but instead employ a marginalization technique to handle this nuisance parameter 
when imposing cosmological constraints~\cite{Conley2011SUPERNOVACA,Bouali:2019whr,Gong:2007se,PhysRevD.72.123519,2001A&A...380....6G}. 
The $\chi^{2}$ statistic for the marginalized method applied 
to SNIa data can be formulated as follows~\cite{Bouali:2019whr,Conley2011SUPERNOVACA}:

\beqa
\chi^{2}_{\rm SNIa}=A-\frac{B^{2}}{C}+{\rm ln}\frac{C}{2\pi},
\eeqa

where 

\beqa
&&A = \Delta \mu^{T}C^{-1}_{\rm SNIa}\Delta \mu \nonumber \\
&&B = \Delta \mu^{T}C^{-1}_{\rm SNIa}I \nonumber \\
&&C = I^{T}C^{-1}_{\rm SNIa}I
\eeqa
Here, $\Delta \mu = \mu_{\rm th}-\mu_{\rm obs}$, $I$ denotes the identity matrix, and 
$C_{\rm SNIa}$ is the covariance matrix of the SNIa data, encompassing statistical and systematic uncertainties. 

For our analysis, we utilize the following supernova (SNIa) datasets: 
(i) the PantheonPlus sample~\footnote{\url{https://github.com/PantheonPlusSH0ES/DataRelease}}, 
comprising 1701 light curves from 1550 distinct SNIa, spanning a redshift range of $0.001<z<2.26$~\cite{Scolnic:2021amr,Brout:2022vxf}. 
To mitigate the effects of peculiar velocities in nearby galaxies, 
we restrict our analysis to data with $z>0.01$; 
(ii) the five-year SNIa data from the Dark Energy Survey (DES)~\footnote{\url{https://github.com/des-science/DES-SN5YR}}, 
which includes 1635 SNIa covering a redshift range of $0.1<z<1.3$, 
supplemented by an additional 194 low-redshift SNIa in the range $0.025<z<0.1$. 
This combined dataset, totaling 1829 SNIa, is commonly referred to as 
DESY5~\cite{DES:2024jxu,2009ApJ...700..331H,2012ApJS..200...12H,Krisciunas:2017yoe,Foley:2017zdq}. 
(iii) the Union3 sample, consisting of 2087 SNIa~\cite{Rubin:2023jdq}. For our calculations, 
we employ the binned data~\footnote{\url{https://github.com/CobayaSampler/sn_data}}, 
which spans $0.05<z<2.26$. We have verified the consistency of our constraints—for instance, in the 
$w_{0}w_{a}\rm CDM$ cosmological model—with previous results, such as those reported in Ref.~\cite{DESI:2025zgx}.

Note that the effects of deviations from $\Lambda$CDM on large-scale structure, which have been explored, for example, 
in Refs.~\cite{8ync-vrtz,SolaPeracaula:2017esw,Nesseris:2017vor,Khatri:2024yfr,Sahlu:2023wvl,Kazantzidis:2018rnb}, 
are not included in this work. We leave this issue for future study.
 
%%%%%%%%%%%%%%%%%%%%%%% SN Ia  %%%%%%%%%%%%%%%%%%%%%%%%%%%%%%%%%%

\section{Constraints on considered models}
\label{cons}
\begin{figure*}[htbp]
\centering
\begin{minipage}[t]{0.48\textwidth}
\centering
\includegraphics[width=8.5cm]{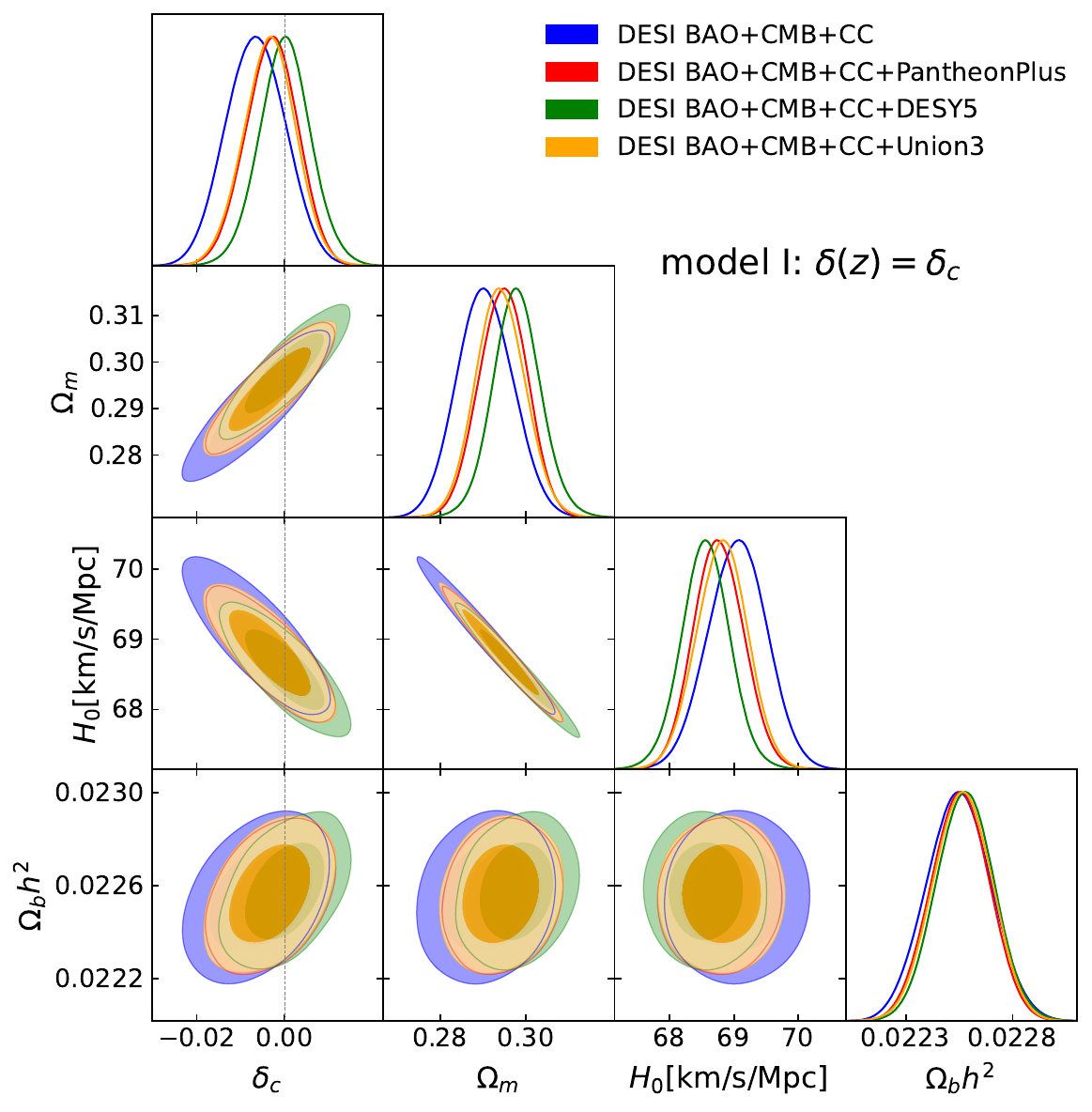}
\end{minipage}

\begin{minipage}[t]{0.48\textwidth}
\centering
\includegraphics[width=8.5cm]{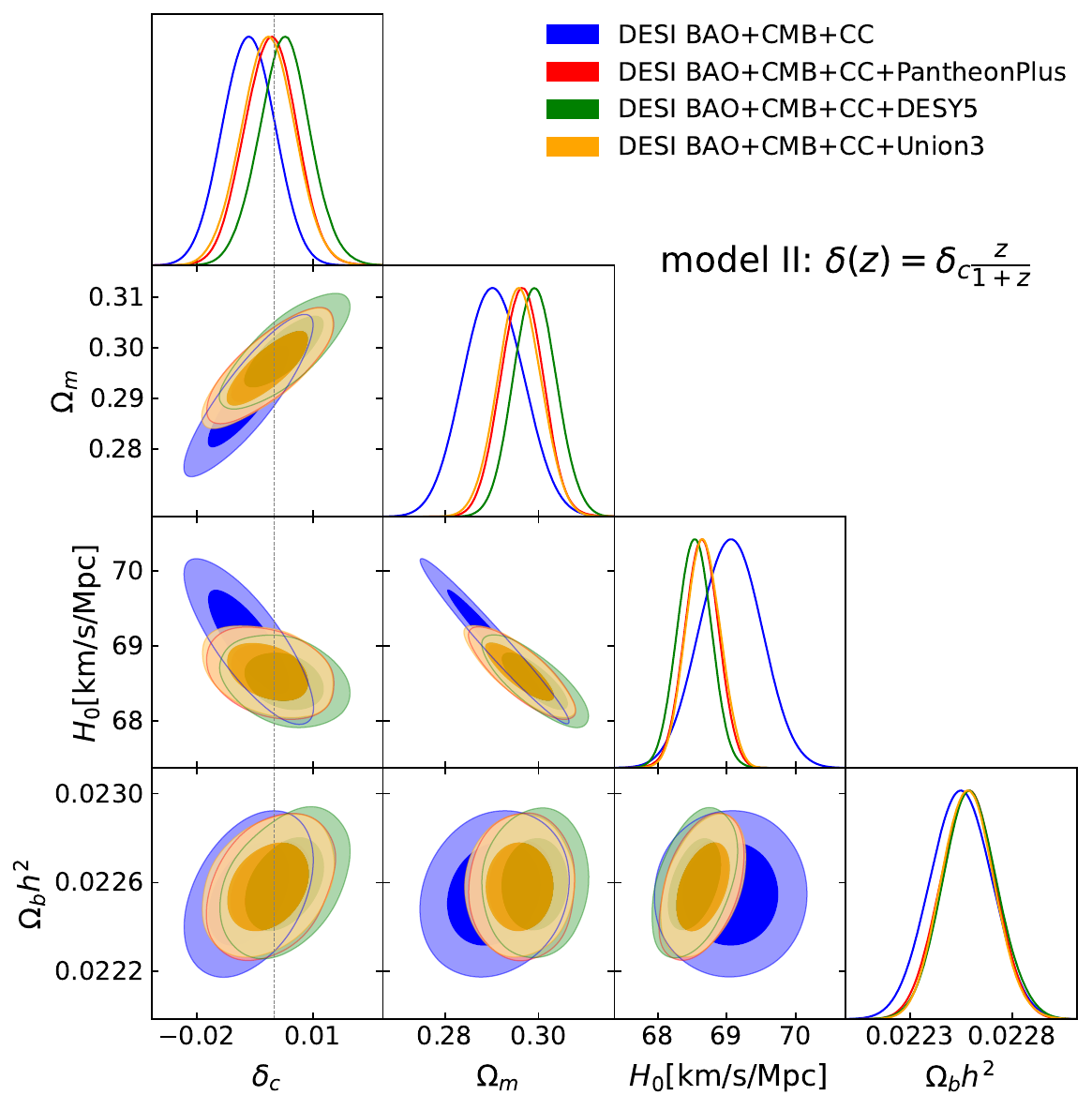}

\end{minipage}
\begin{minipage}[t]{0.48\textwidth}
\centering
\includegraphics[width=8.5cm]{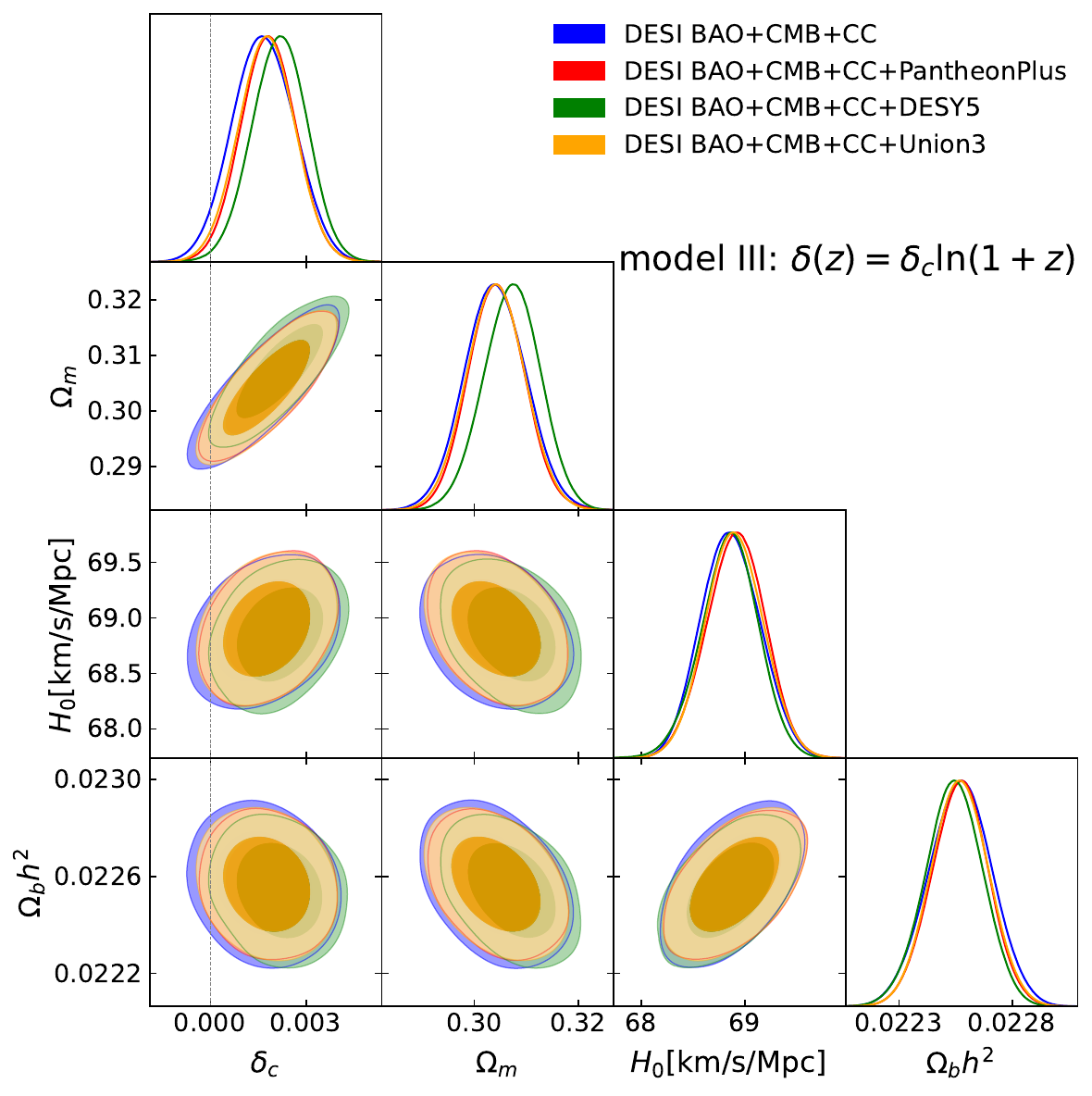}

\end{minipage}
\caption{One-dimensional marginalized posterior distributions and two-dimensional confidence contours (68\% and 95\% C.L.) 
for the three models, constrained by DESI BAO + CMB + CC (blue), 
and DESI BAO + CMB + CC + PantheonPlus (red)/DESY5 (green)/Union3 (orange). The vertical dashed line in the plots indicates the reference value of 
$\delta_{c} = 0$, which corresponds to the standard $\Lambda$CDM model.}
\label{fig:model}
\end{figure*}

We employ the Markov Chain Monte Carlo (MCMC) method to determined the mean values and posterior distributions of 
the parameter set \{$\delta_c$, $\Omega_{m}$, 
$H_0$, $\Omega_{b}h^2$\}. Given the datasets described in the previous section, the total $\chi^{2}$ is given by 

\beqa
\chi^{2}_{\rm total}=\chi^{2}_{\rm DESI~BAO} + \chi^{2}_{\rm CMB} + \chi^{2}_{\rm CC} + \chi^{2}_{\rm SNIa}
\eeqa
where the likelihood function is $L\propto e^{-\chi^{2}_{\rm total}/2}$. We perform the MCMC sampling using the 
public code $\mathtt{emcee}$~\cite{emcee}, with uniform priors on the parameters: 
$\delta_{c} \in (-1,1)$, $\Omega_{m}\in (0,1)$, $H_{0}\in (50,90)$ and $\Omega_{b}h^{2} \in (0.0001,0.1)$. 
The MCMC chains are analyzed using $\mathtt{GetDist}$~\cite{getdist}, and the mean values of the parameters 
with their $1\sigma$ uncertainties are summarized in Tab.~\ref{table:cons}. 
Fig.~\ref{fig:model} displays the one-dimensional marginalized posterior distributions and 
two-dimensional confidence contours for the parameters, derived from the different dataset combinations: 
DESI BAO + CMB + CC, and DESI BAO + CMB + CC + PantheonPlus/DESY5/Union3 datasets, for three considered models.

%%++++++++++++++++++++++++++ figure 1  and 2 ++++++++++++++++++++++++++++++++++++++

%+++++++++++++++++++++++++++ figure  1 and 2++++++++++++++++++++++++++++++++++++++

%****++++++++++++++++++++++++ Table I ********************************************

\begin{table*}[htb]
\caption{
  {  Parameter constraints (mean values along with $1\sigma$ uncertainties) on the parameter set $\{\delta_c, \Omega_m, H_0, \Omega_b h^2\}$ for 
    the three cosmological models under investigation are presented. 
    The results are derived from various combinations of observational data, namely DESI BAO + CMB + CC 
    and DESI BAO + CMB + CC + PantheonPlus/DESY5/Union3. 
    Entries highlighted in \textcolor{blue}{blue} correspond to cases where the constraint on $\delta_c$ deviates from 
    the $\Lambda$CDM value ($\delta_c=0$) with a significance exceeding $2 \sigma$. }
    Additionally, the differences in the Akaike Information Criterion, defined as 
    $\Delta \rm AIC = AIC_{\rm model} - AIC_{\Lambda CDM}$, 
    for model comparison are also shown, with detailed explanations provided in the main text. 
}
\label{tab_ann}
\begin{center}     
\scriptsize
\begin{tabular}{l|ccccc}
\hline \hline
Model/Dataset & $\delta_c$& $\Omega_{m}$ & $H_0\rm [km/s/Mpc]$ &$\Omega_{b}h^{2}$ &$\Delta \rm AIC$\\

\noalign{\smallskip}\hline\noalign{\smallskip}
Model I~~~~~~~$\delta(z)=\delta_c$ \\
\\
DESI BAO+CMB+CC   &$-0.0065\pm{0.0067}$ &$0.2904\pm{0.0065}$  &$69.05\pm{0.45}$ &$0.02255\pm{0.00015}$ & +1.04  \\        
             
DESI BAO+CMB+CC+PantheonPlus  &$-0.0030\pm{0.0058}$ &$0.2946\pm{0.0056}$  &$68.77\pm{0.38}$ &$0.02255\pm{0.00013}$ & +1.74 \\

DESI BAO+CMB+CC+DESY5  &$\rule{0.8em}{0pt}0.0001\pm{0.0058}$ &$0.2978\pm{0.0056}$  &$68.56\pm{0.37}$ &$0.02258\pm{0.00013}$& +2.01  \\

DESI BAO+CMB+CC+Union3  &$-0.0034\pm{0.0059}$ &$0.2940\pm{0.0057}$  &$68.81\pm{0.37}$ &$0.02256\pm{0.00013}$& +1.82 \\
\noalign{\smallskip}\hline\noalign{\smallskip}
Model II~~~~~~ $\delta(z)=\delta_c z/(1+z)$\\
\\
DESI BAO+CMB+CC   &$-0.0054\pm{0.0077}$ &$0.2931\pm{0.0052}$  &$68.76\pm{0.28}$ &$0.02257\pm{0.00015}$ & +1.65  \\             
             
DESI BAO+CMB+CC+PantheonPlus  &$-0.0009\pm{0.0066}$ &$0.2964\pm{0.0045}$  &$68.64\pm{0.24}$ &$0.02258\pm{0.00013}$ & +1.98 \\

DESI BAO+CMB+CC+DESY5  &$\rule{0.8em}{0pt}0.0027\pm{0.0066}$ &$0.2992\pm{0.0045}$  &$68.53\pm{0.24}$ &$0.02259\pm{0.00013}$& +1.86  \\

DESI BAO+CMB+CC+Union3  &$-0.0016\pm{0.0068}$ &$0.2959\pm{0.0046}$  &$68.66\pm{0.25}$ &$0.02258\pm{0.00013}$ & +2.14 \\

\noalign{\smallskip}\hline\noalign{\smallskip}
Model III~~~~~~$\delta(z)=\delta_c{\rm ln}(1+z)$ \\
\\
DESI BAO+CMB+CC   &$\rule{0.8em}{0pt}0.00167\pm{0.00096}$&$0.3041\pm{0.0058}$  &$68.86\pm{0.28}$ &$0.02257\pm{0.00014}$ & $-1.04$  \\             
             
DESI BAO+CMB+CC+PantheonPlus  &\textcolor{blue}{$\rule{0.8em}{0pt}0.00181\pm{0.00085}$} &$0.3043\pm{0.0053}$  &$68.91\pm{0.28}$ &$0.02257\pm{0.00012}$ & $-9.80$ \\

DESI BAO+CMB+CC+DESY5  &\textcolor{blue}{$\rule{0.8em}{0pt}0.00217\pm{0.00086}$} &$0.3073\pm{0.0055}$  &$68.85\pm{0.26}$ &$0.02258\pm{0.00012}$& $-11.63$  \\

DESI BAO+CMB+CC+Union3  &\textcolor{blue}{$\rule{0.8em}{0pt}0.00176\pm{0.00087}$}  &$0.3042\pm{0.0055}$  &$68.90\pm{0.27}$ &$0.02257\pm{0.00012}$ & $-9.41$ \\

\hline \hline
\end{tabular}

\end{center}
\label{table:cons}
\end{table*}

%+++++++++++++++++++++++ Table I ****************************************

For model I, which features a redshift-independent deviation parameter $\delta(z)=\delta_c$, 
the mean value of $\delta_c$ is predominantly negative across all combined datasets, 
except for DESI BAO + CMB + CC + DESY5. For all combined datasets, the deviation from the standard $\Lambda$CDM scenario 
is less than 1$\sigma$, indicating the agreement with the standard model. 
Additionally, as shown in Fig.~\ref{fig:model}, the degeneracy direction among the parameters 
remains nearly identical across all datasets for model I.

For model II, where the deviation parameter $\delta(z) = \delta_c z / (1+z)$ is redshift-dependent, 
the mean value of $\delta_c$  is also mostly negative across the combined datasets, again except for DESI BAO + CMB + CC + DESY5. 
In this model, the discrepancy from $\Lambda$CDM remains below $1\sigma$ 
significance, indicating consistency with the standard cosmological scenario. 
Unlike model I, the degeneracy direction among the parameters in model II exhibits notable variations upon 
the inclusion of SNIa datasets, as illustrated in Fig.~\ref{fig:model}. 
The model exhibits characteristic redshift-dependent behavior: at low redshifts ($z\ll 1$), $\delta(z)$ scales approximately 
linearly as $\delta_{c}z$, while at high redshifts ($z\gg 1$) it asymptotically approaches $\delta_c$. 
This distinctive redshift dependence should help explain why model II, unlike model I, shows significant variations 
in parameter degeneracy directions when SNIa datasets are included. Since current SNIa samples primarily probe 
the low-redshift universe, they mainly constrain the linear regime of the deviation parameter.

For model III, which incorporates a redshift-independent deviation parameter $\delta(z) = \delta_{c} {\rm ln}(1+z)$, 
the mean value of $\delta_c$ remains consistently positive across all combined datasets. This model exhibits a 
particularly compelling feature: compared to the standard $\Lambda$CDM model (where $\delta_{c} = 0$), 
the inferred $\delta_c$ values show statistically significant deviations for most datasets. 
The analysis reveals notable departures from the standard model, with deviations reaching $\sim 1.7\sigma$ for the 
DESI BAO + CMB + CC dataset combination. The inclusion of SNIa datasets 
further enhances these deviations, yielding $\sim 2.1/2.0 \sigma$ for DESI BAO + CMB + CC + PantheonPlus/Union3 and 
$\sim 2.5\sigma$ for DESI BAO + CMB + CC + DESY5. 
In this model, the deviation parameter follows $\delta(z)\sim \delta_{c}z$ at low redshifts ($z<1$), 
exhibiting behavior qualitatively similar to model II. However, a more detailed comparison of the functional forms, 
specifically $z/(1+z)$ for model II versus $\ln(1+z)$ for model III, reveals important distinctions. 
The two functions show relatively minor differences (approximately 10\%) in the low-redshift regime ($z<0.2$). 
However, their divergence becomes increasingly pronounced at higher redshifts ($z>0.2$), growing systematically 
with redshift. These functional differences should lead to distinct parameter degeneracy directions between models II and III, 
which may have significant implications for cosmological parameter estimation. 

At high redshifts, model I asymptotically approaches model II, which should provide explains their similar parameter degeneracy 
signatures. In contrast, model III exhibits fundamentally different behavior from models I and II in this regime. 
This distinction should have contributions to the reversed degeneracy directions between the models. 
A particularly revealing example is the correlation between $\delta_c$ and $H_0$ as shown in Fig.~\ref{fig:model}: 
while these parameters show negative degeneracy in both models I and II, their relationship becomes positive in model III. 
This sign reversal in parameter correlations underscores the unique characteristics of model III's functional form at high redshifts. 
Moreover, the different behave of model III from models I and II at high redshift should also result in 
the prefered positive mean values of $\delta_c$ for model III compared with models I and II, where almost all the 
mean values of $\delta_c$ are negative.

%%%%%%%%%%%%%%%  AIC  %%%%%%%%%%%%%%%%%%%% 
To facilitate the comparison of different models, we integrate model selection criteria through the use of 
the Akaike Information Criterion (AIC)~\cite{Akaike1974A,1978AnSta...6..461S,Kass01061995}, which is mathematically defined 
as follows:

\beqa
{\rm AIC} = \chi^{2}_{\rm min} + \frac{2nN}{N-n-1}
\eeqa
Here, $n$ represents the number of free parameters in the model, and $N$ denotes the total number of data points 
utilized for the cosmological model under consideration. To assess the relative preference between a given model and the benchmark 
$\Lambda$CDM model using observational datasets, we compute the difference $\Delta {\rm AIC} = {\rm AIC_{model}} - {\rm AIC_{\Lambda CDM}}$, with $\Delta \rm AIC_{\rm \Lambda CDM} \equiv 0$ by definition. The interpretation of $|\Delta \rm AIC|$ values follows the established criteria~\cite{SolaPeracaula:2017esw}:

\begin{itemize}
\item[$\bullet$] $|\Delta \rm AIC|<2$: The models are statistically equivalent
\item[$\bullet$] $2<|\Delta \rm AIC|<6$: The model with higher AIC has marginally less support
\item[$\bullet$] $6<|\Delta \rm AIC|<10$: The model with higher AIC is substantially disfavored
\item[$\bullet$] $|\Delta \rm AIC|>10$: The model with higher AIC is strongly ruled out, with compelling evidence favoring the alternative
\end{itemize}

The $\Delta \rm AIC$ values of different models and combined datasets are presented in Tab.~\ref{table:cons}. 
For model I, all $\Delta \rm AIC$ values are positive, with most being less than 2 (+2.01 for the combination of 
DESI BAO + CMB + CC + DESY5). 
This indicates that the model I is statistically 
equivalent to the $\Lambda$CDM model. 
For model II, all $\Delta \rm AIC$ values are also positive, and most fall below 2 (+2.14 for the combination of 
DESI BAO + CMB + CC + Union3). 
This demonstrates that model II is either statistically equivalent to the $\Lambda$CDM model or receives less statistical support. 
For model III, the  $\Delta \rm AIC$ values for different combined datasets are all negative. 
For the datasets DESI BAO + CMB + CC 
the $\Delta \rm AIC$ value ($-1.04$) implies that this model is statistically equivalent to the $\Lambda$CDM model. 
For other datasets, such as DESI BAO + CMB + CC + PantheonPlus and DESI BAO + CMB + CC + Union3, 
the $\Delta \rm AIC$ values ($-9.80$ and $-9.41$) fall winthin the range of $6<|\Delta \rm AIC|<10$, indicating that the $\Lambda$CDM model 
has substantially disfavored 
compared with model III. For the combined datasets DESI BAO + CMB + CC + DESY5, 
the $\Delta \rm AIC=-11.61$ falls within the range of $|\Delta \rm AIC|>10$, 
suggesting that the $\Lambda$CDM model is disfavored when compared with model III. 
In summary, based on the AIC, for the considered models and 
full datasets DESI + CMB + CC + SNIa, model III is preferred over the standard $\Lambda$CDM model.

Based on the constraints obtained from the three considered models and the utilized datasets, 
we find a significant deviation in the Hubble expansion rate from the standard $\Lambda$CDM model for the parameterization 
$\delta(z) = \delta_{c} {\rm ln}(1+z)$ (model III). If the true Hubble expansion rate of the Universe strictly follows the $\Lambda$CDM paradigm, 
no significant deviation should be detected for any combination of datasets or any parameterization form 
(assuming, of course, that the observational data are correct, as we have adopted in this analysis). 
While models I and II show no statistically significant deviation from $\Lambda$CDM for nearly all datasets, 
the pronounced deviation exhibited by model III suggests that the Hubble expansion rate may indeed depart from the 
$\Lambda$CDM prediction. This conclusion is further supported by the Akaike Information Criterion (AIC) values, 
which favor model III over the standard $\Lambda$CDM framework. 

However, despite the statistical preference for model III 
and its significant deviation from $\Lambda$CDM, we caution against interpreting this as evidence that model III represents 
the true Hubble expansion rate of the Universe. Instead, it may simply indicate one possible expansion scenario among many. 
This situation is analogous to the ongoing debate surrounding dark energy. As discussed in Refs.~\cite{Samsing_2012,Hojjati:2013oya}, 
the deviation $\delta(z)$ can be equivalently interpreted as a manifestation of dynamic dark energy rather than a cosmological constant. 
Thus, different functional forms of $\delta(z)$ effectively correspond to different dynamical dark energy models. 
While numerous observational datasets favor dynamical dark energy models (e.g., the CPL parameterization) over a pure cosmological constant, 
the true nature of dark energy remains elusive. Future detailed researches on this issue would be meaningful.

%%%%%%%%%%%%%%%%%%%%%%%%%%%%%%%%%%%% The potential cosmological model %%%%%%%%%%%%%%%%%%%

\section{The potential cosmological model}
\label{model}
 
In view of the final constrains on the three models, it can be seen that the model III is preferred by the 
cosmological datasets compared with the other two models. One question is what kind of cosmological model 
lies behind model III, or what model(s) exhibit similar behavior to model III. 
We here provide a brief discussions on this issue, while a detailed investigation is beyond the scope of 
the current work and will be left for future studies. 

For the convenience of discussion, we consider the low-redshift epoch where the matter dominates and 
ignore the contribution of radiation. For simplicity, we also exclude the baryonic matter and set $\delta_{c}>0$. 
It is worth noting that when $\delta_{c}\ll 1$, the term $[1+\delta_{c}{\rm ln}(1+z)]$ (model III) asymptotically approaches $(1+z)^{\delta_c}$.  
For model III investigated in this work, Eq.~(\ref{eq:basic}) can thus be rewritten as follows,

\beqa
&&H^{2}=H^{2}_{0}\left[\Omega_{m}(1+z)^{3+\delta_c}+\Omega_{\Lambda}(1+z)^{\delta_c}\right]
\label{eq:basic_other}
\eeqa 

Interestingly, for the Hubble expansion rate expressed in the form of Eq.~(\ref{eq:basic_other}), it can be shown that 
interacting cosmological models involving dark energy and dark matter with a constant coupling parameter exhibit analogous behavior~\cite{Cai:2004dk}. 
Below, we outline the key features of this interacting model, and one can refer to Ref.~\cite{Cai:2004dk} for a more comprehensive discussion. 

For the interacting cosmological model, the continuity equations in a flat Friedmann-Lema\^{i}tre-Robertson-Walker (FLRW) metric can be written as,

\beqa
&&\dot{\rho}_{m}+3H\rho_{m}=-Q %\delta_{c}H\rho_{m}
\label{eq:inter1}\\
&& \dot{\rho}_{de}+3H\rho_{de}(1+w_{de})=Q%\delta_{c}H\rho_{m}
\label{eq:inter2}
\eeqa 
Here, $Q$ denotes the interaction between the dark energy and dark matter, and many functional forms for $Q$ 
has been proposed (see, e.g., Ref.~\cite{Wang:2016lxa} for a review). 
For the purposes of this work, we adopt the interaction term $Q=\delta_{c}H\rho_{m}$. The parameter $\delta_c$ is consistent with that 
used in model III of this work, and here it denotes a constant describing the coupling 
between the dark energy and dark matter. 
$\rho_{m}$ ($\rho_{de}$) represents the energy density of dark matter (dark energy), and $w_{de}$ is the equation of state parameter of dark energy. 
From Eq.~(\ref{eq:inter1}), the energy density of dark matter can be derived as,

\beqa
\rho_{m}=\rho_{m_0}(1+z)^{3+\delta_c}
\label{eq:rho_m}
\eeqa

Following the model proposed in Refs.~\cite{Cai:2004dk,Majerotto:2004ji,Dalal:2001dt}, 
we assume that the energy densities of dark energy and dark matter satisfy the scaling relation, 

\beqa
\frac{\rho_{de}}{\rho_m}=\frac{\rho_{de_0}}{\rho_{m_0}}\left(\frac{1}{1+z}\right)^{\varepsilon}
\eeqa
where $\rho_{m_0}$ ($\rho_{de_0}$) is the energy density of dark matter (dark energy) at redshift $z=0$. 
By combining this relation with Eq.~(\ref{eq:rho_m}), the energy density of dark energy can be expressed as, 

\beqa
\rho_{de}=\rho_{de_0}(1+z)^{3+\delta_c-\varepsilon}
\label{eq:rho_de}
\eeqa

Subsequently, the Hubble expansion rate $H_{\rm int}$ for this interacting cosmological model takes the form, 

\beqa
H^{2}_{\rm int}=H^{2}_{0}\left[\Omega_{m_0}(1+z)^{3+\delta_c}+\Omega_{de_0}(1+z)^{3+\delta_c-\varepsilon}\right]
\label{eq:hubble_expansion_rate}
\eeqa

It is evident from the above equation that if we set $\varepsilon=3$, Eq.~(\ref{eq:hubble_expansion_rate}) 
becomes formally identical to Eq.~(\ref{eq:basic_other}).

%%%%%%%%%%%%%%%%%%%%%%%%%%%%%%%%%%%%%%%%%%%  conclusions %%%%%%%%%%%%%%%%%%%%%%%%%%
\section{Conclusions}
\label{con}

We have constrained three models of the Hubble expansion rate using datasets from DESI baryon acoustic oscillation (DESI BAO), 
cosmic microwave background (CMB), cosmic chronometers (CC), 
and Type Ia supernovae (PantheonPlus, DESY5, and Union3). These models describe deviations 
from the standard $\Lambda$CDM model through the parameterization $H^{2}(z)=H^{2}_{\rm \Lambda CDM}(1+\delta(z))$, 
where $\delta(z)$ takes three forms: $\delta(z)=\delta_{c}$ (model I), $\delta_{c}z/(1+z)$ (model II), 
and $\delta_{c}{\rm ln}(1+z)$ (model III), with $\delta_c$ being a constant parameter. 

For most combined datasets, models I and II show no significant deviation from the standard$\Lambda$CDM model 
(i.e., $\delta_{c}=0$ is consistent). 
In contrast to models I and II, model III demonstrates significant deviations 
from the $\Lambda$CDM paradigm for most dataset 
combinations. These deviations reach $\sim 2.1/2.0\sigma$ for DESI BAO + CMB + CC + PantheonPlus/Union3 and 
$\sim 2.5\sigma$ significance when including DESY5 data. 

We performed model comparison using the Akaike Information Criterion (AIC), calculating $\Delta \rm AIC = AIC_{\rm model} - 
AIC_{\Lambda \rm CDM}$ for different models and dataset combinations. The results show that: 
models I and II show statistical equivalence to $\Lambda$CDM across all datasets, 
model III is preferred over $\Lambda$CDM for most combinations (DESI BAO + CMB + CC + SNIa). 
Notably, the $\Delta \rm AIC$ value of $-11.61$ for DESI BAO + CMB + CC + DESY5 strongly 
disfavors $\Lambda$CDM (higher AIC), providing substantial evidence for deviation from the $\Lambda$CDM paradigm 
in the Hubble expansion rate.

In light of recent advancements in cosmology, there is compelling evidence pointing towards dynamic dark energy, 
as opposed to a static cosmological constant. This suggests deviations in the Hubble expansion rate from the 
predictions of the $\Lambda$CDM model. Based on the findings of this study, although model III provides evidence 
of such a deviation, thereby supporting the notion of non-cosmological constant dark energy, 
it would be premature to conclude that model III accurately represents the true Hubble expansion rate. 
Instead, model III offers valuable insights into the potential pathways for the Hubble expansion rate. 
Moreover, we found that some interacting cosmological models between dark energy and dark matter 
exhibit behavior similar to that of model III considered in this work. 
Similar to the approaches employed in investigations of the equation of state of dark energy ($w(z)$), 
the form of deviation, $\delta(z)$, can be explored in a model-independent manner, 
as demonstrated in Ref.~\cite{Samsing_2012,Hojjati:2013oya}. Given recent advances in various cosmological observations, 
and since the effects of deviations from $\Lambda$CDM on large-scale structures have not been included in this work, 
a more thorough investigation of this issue remains an important objective for future research.

\section*{Acknowledgements}
We would like to thank the anonymous referee for the constructive and insightful comments and suggestions. We also appreciate the editor's work. 
This work is supported by the Shandong Provincial Natural Science Foundation (Grant No. ZR2025MS16). 
%\

%\newpage 
\newcommand{\bibcommenthead}{}
\bibliographystyle{sn-aps}
\bibliography{ref}

@article{Notari:2024zmi,
    author = "Notari, Alessio and Redi, Michele and Tesi, Andrea",
    title = "{BAO vs. SN evidence for evolving dark energy}",
    eprint = "2411.11685",
    archivePrefix = "arXiv",
    primaryClass = "astro-ph.CO",
    doi = "10.1088/1475-7516/2025/04/048",
    journal = "JCAP",
    volume = "04",
    pages = "048",
    year = "2025"
}

@article{Dalal:2001dt,
    author = "Dalal, Neal and Abazajian, Kevork and Jenkins, Elizabeth Ellen and Manohar, Aneesh V.",
    title = "{Testing the cosmic coincidence problem and the nature of dark energy}",
    eprint = "astro-ph/0105317",
    archivePrefix = "arXiv",
    doi = "10.1103/PhysRevLett.87.141302",
    journal = "Phys. Rev. Lett.",
    volume = "87",
    pages = "141302",
    year = "2001"
}

@article{Gialamas:2024lyw,
    author = {Gialamas, Ioannis D. and H{\"u}tsi, Gert and Kannike, Kristjan and Racioppi, Antonio and Raidal, Martti and Vasar, Martin and Veerm{\"a}e, Hardi},
    title = "{Interpreting DESI 2024 BAO: Late-time dynamical dark energy or a local effect?}",
    eprint = "2406.07533",
    archivePrefix = "arXiv",
    primaryClass = "astro-ph.CO",
    doi = "10.1103/PhysRevD.111.043540",
    journal = "Phys. Rev. D",
    volume = "111",
    number = "4",
    pages = "043540",
    year = "2025"
}

@article{Wang:2007mza,
    author = "Wang, Yun and Mukherjee, Pia",
    title = "{Observational Constraints on Dark Energy and Cosmic Curvature}",
    eprint = "astro-ph/0703780",
    archivePrefix = "arXiv",
    doi = "10.1103/PhysRevD.76.103533",
    journal = "Phys. Rev. D",
    volume = "76",
    pages = "103533",
    year = "2007"
}

@article{Zhai:2019nad,
    author = "Zhai, Zhongxu and Park, Chan-Gyung and Wang, Yun and Ratra, Bharat",
    title = "{CMB distance priors revisited: effects of dark energy dynamics, spatial curvature, primordial power spectrum, and neutrino parameters}",
    eprint = "1912.04921",
    archivePrefix = "arXiv",
    primaryClass = "astro-ph.CO",
    doi = "10.1088/1475-7516/2020/07/009",
    journal = "JCAP",
    volume = "07",
    pages = "009",
    year = "2020"
}

@article{Efstathiou:2025tie,
    author = "Efstathiou, George",
    title = "{Baryon Acoustic Oscillations from a Different Angle}",
    eprint = "2505.02658",
    archivePrefix = "arXiv",
    primaryClass = "astro-ph.CO",
    month = "5",
    year = "2025"
}

@article{Huang:2025som,
    author = "Huang, Lu and Cai, Rong-Gen and Wang, Shao-Jiang",
    title = "{The DESI DR1/DR2 evidence for dynamical dark energy is biased by low-redshift supernovae}",
    eprint = "2502.04212",
    archivePrefix = "arXiv",
    primaryClass = "astro-ph.CO",
    doi = "10.1007/s11433-025-2754-5",
    journal = "Sci. China Phys. Mech. Astron.",
    volume = "68",
    number = "10",
    pages = "100413",
    year = "2025"
}

@article{Li:2024qso,
    author = "Li, Tian-Nuo and Wu, Peng-Ju and Du, Guo-Hong and Jin, Shang-Jie and Li, Hai-Li and Zhang, Jing-Fei and Zhang, Xin",
    title = "{Constraints on Interacting Dark Energy Models from the DESI Baryon Acoustic Oscillation and DES Supernovae Data}",
    eprint = "2407.14934",
    archivePrefix = "arXiv",
    primaryClass = "astro-ph.CO",
    doi = "10.3847/1538-4357/ad87f0",
    journal = "Astrophys. J.",
    volume = "976",
    number = "1",
    pages = "1",
    year = "2024"
}

@article{Adhikary:2025khr,
    author = "Adhikary, Priyanka and Das, Sudipta and Odintsov, Sergei D. and Paul, Tanmoy",
    title = "{Dark energy era with a resolution of Hubble tension in generalized entropic cosmology}",
    eprint = "2507.15273",
    archivePrefix = "arXiv",
    primaryClass = "gr-qc",
    doi = "10.1016/j.dark.2025.102037",
    journal = "Phys. Dark Univ.",
    volume = "49",
    pages = "102037",
    year = "2025"
}

@article{Nojiri:2025low,
    author = "Nojiri, Shin'ichi and Odintsov, S. D. and Oikonomou, V. K.",
    title = "{Phantom Crossing and Oscillating Dark Energy with $F(R)$ Gravity}",
    eprint = "2506.21010",
    archivePrefix = "arXiv",
    primaryClass = "gr-qc",
    reportNumber = "KEK-TH-2734, KEK-Cosmo-0383",
    month = "6",
    year = "2025"
}

@article{Odintsov:2025jfq,
    author = "Odintsov, S. D. and Oikonomou, V. K. and Sharov, G. S.",
    title = "{Dynamical Dark Energy from $F(R)$ Gravity Models Unifying Inflation with Dark Energy: Confronting the Latest Observational Data}",
    eprint = "2506.02245",
    archivePrefix = "arXiv",
    primaryClass = "gr-qc",
    month = "6",
    year = "2025"
}

@article{Odintsov:2024woi,
    author = "Odintsov, Sergei D. and S{\'a}ez-Chill{\'o}n G{\'o}mez, Diego and Sharov, German S.",
    title = "{Modified gravity/dynamical dark energy vs $\Lambda $CDM: is the game over?}",
    eprint = "2412.09409",
    archivePrefix = "arXiv",
    primaryClass = "gr-qc",
    doi = "10.1140/epjc/s10052-025-14013-3",
    journal = "Eur. Phys. J. C",
    volume = "85",
    number = "3",
    pages = "298",
    year = "2025"
}

@article{Braglia:2025gdo,
    author = "Braglia, Matteo and Chen, Xingang and Loeb, Abraham",
    title = "{Exotic Dark Matter and the DESI Anomaly}",
    eprint = "2507.13925",
    archivePrefix = "arXiv",
    primaryClass = "astro-ph.CO",
    month = "7",
    year = "2025"
}

@article{RoyChoudhury:2024wri,
    author = "Roy Choudhury, Shouvik and Okumura, Teppei",
    title = "{Updated Cosmological Constraints in Extended Parameter Space with Planck PR4, DESI Baryon Acoustic Oscillations, and Supernovae: Dynamical Dark Energy, Neutrino Masses, Lensing Anomaly, and the Hubble Tension}",
    eprint = "2409.13022",
    archivePrefix = "arXiv",
    primaryClass = "astro-ph.CO",
    doi = "10.3847/2041-8213/ad8c26",
    journal = "Astrophys. J. Lett.",
    volume = "976",
    number = "1",
    pages = "L11",
    year = "2024"
}

@article{RYSKIN2015258,
title = {The emergence of cosmic repulsion},
journal = {Astroparticle Physics},
volume = {62},
pages = {258-268},
year = {2015},
issn = {0927-6505},
doi = {https://doi.org/10.1016/j.astropartphys.2014.10.003},
url = {https://www.sciencedirect.com/science/article/pii/S0927650514001583},
author = {Gregory Ryskin}
}

@article{Vagnozzi:2019ezj,
    author = "Vagnozzi, Sunny",
    title = "{New physics in light of the $H_0$ tension: An alternative view}",
    eprint = "1907.07569",
    archivePrefix = "arXiv",
    primaryClass = "astro-ph.CO",
    doi = "10.1103/PhysRevD.102.023518",
    journal = "Phys. Rev. D",
    volume = "102",
    number = "2",
    pages = "023518",
    year = "2020"
}

@article{Vagnozzi:2023nrq,
    author = "Vagnozzi, Sunny",
    title = "{Seven Hints That Early-Time New Physics Alone Is Not Sufficient to Solve the Hubble Tension}",
    eprint = "2308.16628",
    archivePrefix = "arXiv",
    primaryClass = "astro-ph.CO",
    doi = "10.3390/universe9090393",
    journal = "Universe",
    volume = "9",
    number = "9",
    pages = "393",
    year = "2023"
}

@article{Colgain:2025nzf,
    author = "Colg{\'a}in, Eoin {\'O}. and Pourojaghi, Saeed and Sheikh-Jabbari, M. M. and Yin, Lu",
    title = "{How much has DESI dark energy evolved since DR1?}",
    eprint = "2504.04417",
    archivePrefix = "arXiv",
    primaryClass = "astro-ph.CO",
    month = "4",
    year = "2025"
}

@article{CosmoVerseNetwork:2025alb,
    author = "Di Valentino, Eleonora and others",
    collaboration = "CosmoVerse Network",
    title = "{The CosmoVerse White Paper: Addressing observational tensions in cosmology with systematics and fundamental physics}",
    eprint = "2504.01669",
    archivePrefix = "arXiv",
    primaryClass = "astro-ph.CO",
    doi = "10.1016/j.dark.2025.101965",
    journal = "Phys. Dark Univ.",
    volume = "49",
    pages = "101965",
    year = "2025"
}

@article{RoyChoudhury:2025dhe,
    author = "Roy Choudhury, Shouvik",
    title = "{Cosmology in Extended Parameter Space with DESI Data Release 2 Baryon Acoustic Oscillations: A 2{\ensuremath{\sigma}}+ Detection of Nonzero Neutrino Masses with an Update on Dynamical Dark Energy and Lensing Anomaly}",
    eprint = "2504.15340",
    archivePrefix = "arXiv",
    primaryClass = "astro-ph.CO",
    doi = "10.3847/2041-8213/ade1cc",
    journal = "Astrophys. J. Lett.",
    volume = "986",
    number = "2",
    pages = "L31",
    year = "2025"
}

@article{Li:2025owk,
    author = "Li, Tian-Nuo and Du, Guo-Hong and Li, Yun-He and Wu, Peng-Ju and Jin, Shang-Jie and Zhang, Jing-Fei and Zhang, Xin",
    title = "{Probing the sign-changeable interaction between dark energy and dark matter with DESI baryon acoustic oscillations and DES supernovae data}",
    eprint = "2501.07361",
    archivePrefix = "arXiv",
    primaryClass = "astro-ph.CO",
    month = "1",
    year = "2025"
}

@article{cpl_1,
author = {CHEVALLIER, MICHEL and POLARSKI, DAVID},
title = {ACCELERATING UNIVERSES WITH SCALING DARK MATTER},
journal = {International Journal of Modern Physics D},
volume = {10},
number = {02},
pages = {213-223},
year = {2001},
doi = {10.1142/S0218271801000822},

URL = { 
    
        https://doi.org/10.1142/S0218271801000822
    
    

},
eprint = { 
    
        https://doi.org/10.1142/S0218271801000822
    
    

}

}

@article{cpl_2,
    author = "Linder, Eric V.",
    title = "{Exploring the expansion history of the universe}",
    eprint = "astro-ph/0208512",
    archivePrefix = "arXiv",
    doi = "10.1103/PhysRevLett.90.091301",
    journal = "Phys. Rev. Lett.",
    volume = "90",
    pages = "091301",
    year = "2003"
}

@article{Tripathi:2016slv,
    author = "Tripathi, Ashutosh and Sangwan, Archana and Jassal, H. K.",
    title = "{Dark energy equation of state parameter and its evolution at low redshift}",
    eprint = "1611.01899",
    archivePrefix = "arXiv",
    primaryClass = "astro-ph.CO",
    doi = "10.1088/1475-7516/2017/06/012",
    journal = "JCAP",
    volume = "06",
    pages = "012",
    year = "2017"
}

@article{Wang:2025znm,
    author = "Wang, Jia-Qi and Cai, Rong-Gen and Guo, Zong-Kuan and Wang, Shao-Jiang",
    title = "{Resolving the Planck-DESI tension by non-minimally coupled quintessence}",
    eprint = "2508.01759",
    archivePrefix = "arXiv",
    primaryClass = "astro-ph.CO",
    month = "8",
    year = "2025"
}

@article{Efstathiou:1999tm,
    author = "Efstathiou, G.",
    title = "{Constraining the equation of state of the universe from distant type Ia supernovae and cosmic microwave background anisotropies}",
    eprint = "astro-ph/9904356",
    archivePrefix = "arXiv",
    doi = "10.1046/j.1365-8711.1999.02997.x",
    journal = "Mon. Not. Roy. Astron. Soc.",
    volume = "310",
    pages = "842--850",
    year = "1999"
}

@article{Moresco:2024wmr,
    author = "Moresco, Michele",
    title = "{Measuring the expansion history of the Universe with cosmic chronometers}",
    eprint = "2412.01994",
    archivePrefix = "arXiv",
    primaryClass = "astro-ph.CO",
    month = "12",
    year = "2024"
}

@article{Moresco:2020fbm,
    author = "Moresco, Michele and Jimenez, Raul and Verde, Licia and Cimatti, Andrea and Pozzetti, Lucia",
    title = "{Setting the Stage for Cosmic Chronometers. II. Impact of Stellar Population Synthesis Models Systematics and Full Covariance Matrix}",
    eprint = "2003.07362",
    archivePrefix = "arXiv",
    primaryClass = "astro-ph.GA",
    doi = "10.3847/1538-4357/ab9eb0",
    journal = "Astrophys. J.",
    volume = "898",
    number = "1",
    pages = "82",
    year = "2020"
}

@article{Silva:2011pc,
    author = "Silva, R. and Goncalves, R. S. and Alcaniz, J. S. and Silva, H. H. B.",
    title = "{Thermodynamics and dark energy}",
    eprint = "1104.1628",
    archivePrefix = "arXiv",
    primaryClass = "astro-ph.CO",
    doi = "10.1051/0004-6361/201117707",
    journal = "Astron. Astrophys.",
    volume = "537",
    pages = "A11",
    year = "2012"
}

@article{Perivolaropoulos:2021jda,
    author = "Perivolaropoulos, Leandros and Skara, Foteini",
    title = "{Challenges for \ensuremath{\Lambda}CDM: An update}",
    eprint = "2105.05208",
    archivePrefix = "arXiv",
    primaryClass = "astro-ph.CO",
    doi = "10.1016/j.newar.2022.101659",
    journal = "New Astron. Rev.",
    volume = "95",
    pages = "101659",
    year = "2022"
}

@article{Bull:2015stt,
    author = "Bull, Philip and others",
    title = "{Beyond $\Lambda$CDM: Problems, solutions, and the road ahead}",
    eprint = "1512.05356",
    archivePrefix = "arXiv",
    primaryClass = "astro-ph.CO",
    doi = "10.1016/j.dark.2016.02.001",
    journal = "Phys. Dark Univ.",
    volume = "12",
    pages = "56--99",
    year = "2016"
}

@article{Lemos:2025qyh,
    author = "Lemos, Thais and Gon{\c{c}}alves, Rodrigo and Carvalho, Joel and Alcaniz, Jailson",
    title = "{Cosmological-model independent limits on photon mass from FRB and SNe data}",
    eprint = "2504.21129",
    archivePrefix = "arXiv",
    primaryClass = "astro-ph.CO",
    month = "4",
    year = "2025"
}

@article{vonMarttens:2025dvv,
    author = "von Marttens, Rodrigo and Gonzalez, Javier and Alcaniz, Jailson",
    title = "{Reconstructing the redshift evolution of Type Ia supernovae absolute magnitude}",
    eprint = "2504.15127",
    archivePrefix = "arXiv",
    primaryClass = "astro-ph.CO",
    month = "4",
    year = "2025"
}

@article{Poulin:2018cxd,
    author = "Poulin, Vivian and Smith, Tristan L. and Karwal, Tanvi and Kamionkowski, Marc",
    title = "{Early Dark Energy Can Resolve The Hubble Tension}",
    eprint = "1811.04083",
    archivePrefix = "arXiv",
    primaryClass = "astro-ph.CO",
    doi = "10.1103/PhysRevLett.122.221301",
    journal = "Phys. Rev. Lett.",
    volume = "122",
    number = "22",
    pages = "221301",
    year = "2019"
}

@article{Wang:2016lxa,
    author = "Wang, B. and Abdalla, E. and Atrio-Barandela, F. and Pavon, D.",
    title = "{Dark Matter and Dark Energy Interactions: Theoretical Challenges, Cosmological Implications and Observational Signatures}",
    eprint = "1603.08299",
    archivePrefix = "arXiv",
    primaryClass = "astro-ph.CO",
    doi = "10.1088/0034-4885/79/9/096901",
    journal = "Rept. Prog. Phys.",
    volume = "79",
    number = "9",
    pages = "096901",
    year = "2016"
}

@article{DESI:2025zgx,
    author = "Abdul Karim, M. and others",
    collaboration = "DESI",
    title = "{DESI DR2 Results II: Measurements of Baryon Acoustic Oscillations and Cosmological Constraints}",
    eprint = "2503.14738",
    archivePrefix = "arXiv",
    primaryClass = "astro-ph.CO",
    reportNumber = "FERMILAB-PUB-25-0169-PPD",
    month = "3",
    year = "2025"
}

@article{DESI:2024mwx,
    author = "Adame, A. G. and others",
    collaboration = "DESI",
    title = "{DESI 2024 VI: Cosmological Constraints from the Measurements of Baryon Acoustic Oscillations}",
    eprint = "2404.03002",
    archivePrefix = "arXiv",
    primaryClass = "astro-ph.CO",
    reportNumber = "FERMILAB-PUB-24-0154-PPD",
    month = "4",
    year = "2024"
}

@article{Feng_2011,
   title={A new equation of state for dark energy model},
   volume={2011},
   ISSN={1475-7516},
   url={http://dx.doi.org/10.1088/1475-7516/2011/11/034},
   DOI={10.1088/1475-7516/2011/11/034},
   number={11},
   journal={Journal of Cosmology and Astroparticle Physics},
   publisher={IOP Publishing},
   author={Feng, Lei and Lu, Tan},
   year={2011},
   month=nov, pages={034–034} }

@article{Akaike1974A,
  title={A new look at the statistical model identification},
  author={Akaike and H.},
  journal={Automatic Control, IEEE Transactions on},
  year={1974},
}

@ARTICLE{1978AnSta...6..461S,
       author = {{Schwarz}, Gideon},
        title = "{Estimating the Dimension of a Model}",
      journal = {Annals of Statistics},
         year = 1978,
        month = jul,
       volume = {6},
       number = {2},
        pages = {461-464},
       adsurl = {https://ui.adsabs.harvard.edu/abs/1978AnSta...6..461S}
}

@article{Kass01061995,
author = {Robert E. Kass and Adrian E. Raftery and},
title = {Bayes Factors},
journal = {Journal of the American Statistical Association},
volume = {90},
number = {430},
pages = {773--795},
year = {1995},
publisher = {ASA Website},
doi = {10.1080/01621459.1995.10476572},

}

@article{emcee,
   title={<tt>emcee</tt>: The MCMC Hammer},
   volume={125},
   ISSN={1538-3873},
   url={http://dx.doi.org/10.1086/670067},
   DOI={10.1086/670067},
   number={925},
   journal={Publications of the Astronomical Society of the Pacific},
   publisher={IOP Publishing},
   author={Foreman-Mackey, Daniel and Hogg, David W. and Lang, Dustin and Goodman, Jonathan},
   year={2013},
   month=mar, pages={306–312} }

@article{getdist,
 author         = "Lewis, Antony",
 title          = "{GetDist: a Python package for analysing Monte Carlo
                   samples}",
 year           = "2019",
 eprint         = "1910.13970",
 archivePrefix  = "arXiv",
 primaryClass   = "astro-ph.IM",
 SLACcitation   = "%%CITATION = ARXIV:1910.13970;%%",
 url            = "https://getdist.readthedocs.io"
}

@article{Xu:2016grp,
    author = "Xu, Yue-Yao and Zhang, Xin",
    title = "{Comparison of dark energy models after Planck 2015}",
    eprint = "1607.06262",
    archivePrefix = "arXiv",
    primaryClass = "astro-ph.CO",
    doi = "10.1140/epjc/s10052-016-4446-5",
    journal = "Eur. Phys. J. C",
    volume = "76",
    number = "11",
    pages = "588",
    year = "2016"
}

@article{Liu:2018kjv,
    author = "Liu, Yan and Guo, Rui-Yun and Zhang, Jing-Fei and Zhang, Xin",
    title = "{Revisit of constraints on dark energy with Hubble parameter measurements including future redshift drift observations}",
    eprint = "1811.12131",
    archivePrefix = "arXiv",
    primaryClass = "astro-ph.CO",
    doi = "10.1088/1475-7516/2019/05/016",
    journal = "JCAP",
    volume = "05",
    pages = "016",
    year = "2019"
}

@article{Hu:2023jqc,
    author = "Hu, Jian-Ping and Wang, Fa-Yin",
    title = "{Hubble Tension: The Evidence of New Physics}",
    eprint = "2302.05709",
    archivePrefix = "arXiv",
    primaryClass = "astro-ph.CO",
    doi = "10.3390/universe9020094",
    journal = "Universe",
    volume = "9",
    number = "2",
    pages = "94",
    year = "2023"
}

@article{DiValentino:2021izs,
    author = "Di Valentino, Eleonora and Mena, Olga and Pan, Supriya and Visinelli, Luca and Yang, Weiqiang and Melchiorri, Alessandro and Mota, David F. and Riess, Adam G. and Silk, Joseph",
    title = "{In the realm of the Hubble tension\textemdash{}a review of solutions}",
    eprint = "2103.01183",
    archivePrefix = "arXiv",
    primaryClass = "astro-ph.CO",
    reportNumber = "IPPP/20/108",
    doi = "10.1088/1361-6382/ac086d",
    journal = "Class. Quant. Grav.",
    volume = "38",
    number = "15",
    pages = "153001",
    year = "2021"
}

@article{8ync-vrtz,
  title = {New cosmological constraints on the evolution of dark matter energy density},
  author = {Yang, Yupeng and Dai, Xinyi and Wang, Yicheng},
  journal = {Phys. Rev. D},
  volume = {111},
  issue = {10},
  pages = {103534},
  numpages = {10},
  year = {2025},
  month = {May},
  publisher = {American Physical Society},
  doi = {10.1103/8ync-vrtz},
  url = {https://link.aps.org/doi/10.1103/8ync-vrtz}
}

@article{Breuval:2024lsv,
    author = "Breuval, Louise and Riess, Adam G. and Casertano, Stefano and Yuan, Wenlong and Macri, Lucas M. and Romaniello, Martino and Murakami, Yukei S. and Scolnic, Daniel and Anand, Gagandeep S. and Soszy\'nski, Igor",
    title = "{Small Magellanic Cloud Cepheids Observed with the Hubble Space Telescope Provide a New Anchor for the SH0ES Distance Ladder}",
    eprint = "2404.08038",
    archivePrefix = "arXiv",
    primaryClass = "astro-ph.CO",
    doi = "10.3847/1538-4357/ad630e",
    journal = "Astrophys. J.",
    volume = "973",
    number = "1",
    pages = "30",
    year = "2024"
}

@ARTICLE{2009ApJ...699..539R,
       author = {{Riess}, Adam G. and {Macri}, Lucas and {Casertano}, Stefano and {Sosey}, Megan and {Lampeitl}, Hubert and {Ferguson}, Henry C. and {Filippenko}, Alexei V. and {Jha}, Saurabh W. and {Li}, Weidong and {Chornock}, Ryan and {Sarkar}, Devdeep},
        title = "{A Redetermination of the Hubble Constant with the Hubble Space Telescope from a Differential Distance Ladder}",
      journal = {apj},
         year = 2009,
        month = jul,
       volume = {699},
       number = {1},
        pages = {539-563},
          doi = {10.1088/0004-637X/699/1/539},
archivePrefix = {arXiv},
       eprint = {0905.0695},
 primaryClass = {astro-ph.CO},
       adsurl = {https://ui.adsabs.harvard.edu/abs/2009ApJ...699..539R}
}

@article{Kang:2019azh,
    author = "Kang, Yijung and Lee, Young-Wook and Kim, Young-Lo and Chung, Chul and Ree, Chang Hee",
    title = "{Early-type Host Galaxies of Type Ia Supernovae. II. Evidence for Luminosity Evolution in Supernova Cosmology}",
    eprint = "1912.04903",
    archivePrefix = "arXiv",
    primaryClass = "astro-ph.GA",
    doi = "10.3847/1538-4357/ab5afc",
    journal = "Astrophys. J.",
    volume = "889",
    number = "1",
    pages = "8",
    year = "2020"
}

@article{Camarena:2021jlr,
    author = "Camarena, David and Marra, Valerio",
    title = "{On the use of the local prior on the absolute magnitude of Type Ia supernovae in cosmological inference}",
    eprint = "2101.08641",
    archivePrefix = "arXiv",
    primaryClass = "astro-ph.CO",
    doi = "10.1093/mnras/stab1200",
    journal = "Mon. Not. Roy. Astron. Soc.",
    volume = "504",
    number = "4",
    pages = "5164--5171",
    year = "2021"
}

@article{Camarena:2019moy,
    author = "Camarena, David and Marra, Valerio",
    title = "{Local determination of the Hubble constant and the deceleration parameter}",
    eprint = "1906.11814",
    archivePrefix = "arXiv",
    primaryClass = "astro-ph.CO",
    doi = "10.1103/PhysRevResearch.2.013028",
    journal = "Phys. Rev. Res.",
    volume = "2",
    number = "1",
    pages = "013028",
    year = "2020"
}

@article{SupernovaCosmologyProject:1998vns,
    author = "Perlmutter, S. and others",
    collaboration = "Supernova Cosmology Project",
    title = "{Measurements of $\Omega$ and $\Lambda$ from 42 High Redshift Supernovae}",
    eprint = "astro-ph/9812133",
    archivePrefix = "arXiv",
    reportNumber = "LBNL-41801, LBL-41801",
    doi = "10.1086/307221",
    journal = "Astrophys. J.",
    volume = "517",
    pages = "565--586",
    year = "1999"
}

@article{SupernovaSearchTeam:1998fmf,
    author = "Riess, Adam G. and others",
    collaboration = "Supernova Search Team",
    title = "{Observational evidence from supernovae for an accelerating universe and a cosmological constant}",
    eprint = "astro-ph/9805201",
    archivePrefix = "arXiv",
    doi = "10.1086/300499",
    journal = "Astron. J.",
    volume = "116",
    pages = "1009--1038",
    year = "1998"
}

@article{Wu:2025vfs,
    author = "Wu, Peng-Ju and Li, Tian-Nuo and Du, Guo-Hong and Zhang, Xin",
    title = "{Observational challenges to holographic and Ricci dark energy paradigms: Insights from ACT DR6 and DESI DR2}",
    eprint = "2509.02945",
    archivePrefix = "arXiv",
    primaryClass = "astro-ph.CO",
    month = "9",
    year = "2025"
}

@article{Zhou:2025nkb,
    author = "Zhou, Sheng-Han and Li, Tian-Nuo and Du, Guo-Hong and Jiang, Jun-Qian and Zhang, Jing-Fei and Zhang, Xin",
    title = "{Measuring neutrino masses with joint JWST and DESI DR2 data}",
    eprint = "2509.10836",
    archivePrefix = "arXiv",
    primaryClass = "astro-ph.CO",
    month = "9",
    year = "2025"
}

@article{Wolf:2025acj,
    author = "Wolf, William J. and Ferreira, Pedro G. and Garc{\'\i}a-Garc{\'\i}a, Carlos",
    title = "{Cosmological constraints on Galileon dark energy with broken shift symmetry}",
    eprint = "2509.17586",
    archivePrefix = "arXiv",
    primaryClass = "astro-ph.CO",
    month = "9",
    year = "2025"
}

@article{RoyChoudhury:2025iis,
    author = "Roy Choudhury, Shouvik and Okumura, Teppei and Umetsu, Keiichi",
    title = "{Cosmological constraints on non-phantom dynamical dark energy with DESI Data Release 2 Baryon Acoustic Oscillations: A 3$\sigma$+ lensing anomaly}",
    eprint = "2509.26144",
    archivePrefix = "arXiv",
    primaryClass = "astro-ph.CO",
    month = "9",
    year = "2025"
}

@article{Alam:2025epg,
    author = "Alam, Sonej and Hossain, Md. Wali",
    title = "{Beyond CPL: Evidence for dynamical dark energy in three-parameter models}",
    eprint = "2510.03779",
    archivePrefix = "arXiv",
    primaryClass = "astro-ph.CO",
    month = "10",
    year = "2025"
}

@article{Li:2025muv,
    author = "Li, Tian-Nuo and Du, Guo-Hong and Li, Yun-He and Li, Yichao and Ling, Jia-Le and Zhang, Jing-Fei and Zhang, Xin",
    title = "{Updated constraints on interacting dark energy: A comprehensive analysis using multiple CMB probes, DESI DR2, and supernovae observations}",
    eprint = "2510.11363",
    archivePrefix = "arXiv",
    primaryClass = "astro-ph.CO",
    month = "10",
    year = "2025"
}

@article{Zhang:2025dwu,
    author = "Zhang, Yi-Min and Li, Tian-Nuo and Du, Guo-Hong and Zhou, Sheng-Han and Gao, Li-Yang and Zhang, Jing-Fei and Zhang, Xin",
    title = "{Alleviating the $H_0$ tension through new interacting dark energy model in light of DESI DR2}",
    eprint = "2510.12627",
    archivePrefix = "arXiv",
    primaryClass = "astro-ph.CO",
    month = "10",
    year = "2025"
}

@article{Cai:2004dk,
    author = "Cai, Rong-Gen and Wang, Anzhong",
    title = "{Cosmology with interaction between phantom dark energy and dark matter and the coincidence problem}",
    eprint = "hep-th/0411025",
    archivePrefix = "arXiv",
    doi = "10.1088/1475-7516/2005/03/002",
    journal = "JCAP",
    volume = "03",
    pages = "002",
    year = "2005"
}

@article{Du:2025csv,
    author = "Du, Guo-Hong and Li, Tian-Nuo and Ling, Jia-Le and Yao, Yan-Hong and Zhang, Jing-Fei and Zhang, Xin",
    title = "{Model-independent late-universe measurements of $H_0$ and $\Omega_\mathrm{K}$ with the PAge-improved inverse distance ladder}",
    eprint = "2510.26355",
    archivePrefix = "arXiv",
    primaryClass = "astro-ph.CO",
    month = "10",
    year = "2025"
}

@article{Majerotto:2004ji,
    author = "Majerotto, Elisabetta and Sapone, Domenico and Amendola, Luca",
    title = "{Supernovae Type Ia data favour negatively coupled phantom energy}",
    eprint = "astro-ph/0410543",
    archivePrefix = "arXiv",
    month = "10",
    year = "2004"
}

@Inbook{Wallisch2019,
author="Wallisch, Benjamin",
title="Review of Modern Cosmology",
bookTitle="Cosmological Probes of Light Relics",
year="2019",
publisher="Springer International Publishing",
address="Cham",
pages="9--47",
isbn="978-3-030-31098-1",
doi="10.1007/978-3-030-31098-1_2",
url="https://doi.org/10.1007/978-3-030-31098-1_2"
}

@article{Brito:2024bhh,
    author = "Brito, L. S. and Jesus, J. F. and Escobal, A. A. and Pereira, S. H.",
    title = "{Can decaying vacuum solve the H\_0 Tension?}",
    eprint = "2412.06756",
    archivePrefix = "arXiv",
    primaryClass = "astro-ph.CO",
    month = "12",
    year = "2024"
}

@article{Hu:1995en,
    author = "Hu, Wayne and Sugiyama, Naoshi",
    title = "{Small scale cosmological perturbations: An Analytic approach}",
    eprint = "astro-ph/9510117",
    archivePrefix = "arXiv",
    reportNumber = "IASSNS-AST-95-42, CFPA-TH-95-18, UTAP-212",
    doi = "10.1086/177989",
    journal = "Astrophys. J.",
    volume = "471",
    pages = "542--570",
    year = "1996"
}

@article{Chen:2018dbv,
    author = "Chen, Lu and Huang, Qing-Guo and Wang, Ke",
    title = "{Distance Priors from Planck Final Release}",
    eprint = "1808.05724",
    archivePrefix = "arXiv",
    primaryClass = "astro-ph.CO",
    doi = "10.1088/1475-7516/2019/02/028",
    journal = "JCAP",
    volume = "02",
    pages = "028",
    year = "2019"
}

@article{Yang:2025vnm,
    author = "Yang, Yupeng and Wang, Yicheng and Dai, Xinyi",
    title = "{Cosmological constraints on two vacuum decay models}",
    eprint = "2502.17792",
    archivePrefix = "arXiv",
    primaryClass = "astro-ph.CO",
    doi = "10.1140/epjc/s10052-025-13990-9",
    journal = "Eur. Phys. J. C",
    volume = "85",
    number = "3",
    pages = "224",
    year = "2025"
}

@article{Yang:2025boq,
    author = "Yang, Yupeng and Dai, Xinyi and Wang, Yicheng",
    title = "{New cosmological constraints on the evolution of dark matter energy density}",
    eprint = "2505.09879",
    archivePrefix = "arXiv",
    primaryClass = "astro-ph.CO",
    doi = "10.1103/8ync-vrtz",
    journal = "Phys. Rev. D",
    volume = "111",
    number = "10",
    pages = "103534",
    year = "2025"
}

@article{Zhai:2018vmm,
    author = "Zhai, Zhongxu and Wang, Yun",
    title = "{Robust and model-independent cosmological constraints from distance measurements}",
    eprint = "1811.07425",
    archivePrefix = "arXiv",
    primaryClass = "astro-ph.CO",
    doi = "10.1088/1475-7516/2019/07/005",
    journal = "JCAP",
    volume = "07",
    pages = "005",
    year = "2019"
}

@ARTICLE{2020A&A...641A...6P,
       author = {Aghanim, N. and others},
        title = "{Planck 2018 results. VI. Cosmological parameters}",
      journal = {A\&A },
         year = 2020,
       volume = {641},
          eid = {A6},
        pages = {A6},
          doi = {10.1051/0004-6361/201833910},
archivePrefix = {arXiv},
       eprint = {1807.06209},
 primaryClass = {astro-ph.CO},
       adsurl = {https://ui.adsabs.harvard.edu/abs/2020A&A...641A...6P}
}

@article{Moresco:2015cya,
    author = "Moresco, Michele",
    title = "{Raising the bar: new constraints on the Hubble parameter with cosmic chronometers at z \ensuremath{\sim} 2}",
    eprint = "1503.01116",
    archivePrefix = "arXiv",
    primaryClass = "astro-ph.CO",
    doi = "10.1093/mnrasl/slv037",
    journal = "Mon. Not. Roy. Astron. Soc.",
    volume = "450",
    number = "1",
    pages = "L16--L20",
    year = "2015"
}

@article{Ratsimbazafy:2017vga,
    author = {Ratsimbazafy, A. L. and Loubser, S. I. and Crawford, S. M. and Cress, C. M. and Bassett, B. A. and Nichol, R. C. and V\"ais\"anen, P.},
    title = "{Age-dating Luminous Red Galaxies observed with the Southern African Large Telescope}",
    eprint = "1702.00418",
    archivePrefix = "arXiv",
    primaryClass = "astro-ph.CO",
    doi = "10.1093/mnras/stx301",
    journal = "Mon. Not. Roy. Astron. Soc.",
    volume = "467",
    number = "3",
    pages = "3239--3254",
    year = "2017"
}

@article{Moresco:2016mzx,
    author = "Moresco, Michele and Pozzetti, Lucia and Cimatti, Andrea and Jimenez, Raul and Maraston, Claudia and Verde, Licia and Thomas, Daniel and Citro, Annalisa and Tojeiro, Rita and Wilkinson, David",
    title = "{A 6\% measurement of the Hubble parameter at $z\sim0.45$: direct evidence of the epoch of cosmic re-acceleration}",
    eprint = "1601.01701",
    archivePrefix = "arXiv",
    primaryClass = "astro-ph.CO",
    doi = "10.1088/1475-7516/2016/05/014",
    journal = "JCAP",
    volume = "05",
    pages = "014",
    year = "2016"
}

@ARTICLE{2012JCAP...08..006M,
       author = {{Moresco}, M. and {Cimatti}, A. and {Jimenez}, R. and {Pozzetti}, L. and {Zamorani}, G. and {Bolzonella}, M. and {Dunlop}, J. and {Lamareille}, F. and {Mignoli}, M. and {Pearce}, H. and {Rosati}, P. and {Stern}, D. and {Verde}, L. and {Zucca}, E. and {Carollo}, C.~M. and {Contini}, T. and {Kneib}, J. -P. and {Le F{\`e}vre}, O. and {Lilly}, S.~J. and {Mainieri}, V. and {Renzini}, A. and {Scodeggio}, M. and {Balestra}, I. and {Gobat}, R. and {McLure}, R. and {Bardelli}, S. and {Bongiorno}, A. and {Caputi}, K. and {Cucciati}, O. and {de la Torre}, S. and {de Ravel}, L. and {Franzetti}, P. and {Garilli}, B. and {Iovino}, A. and {Kampczyk}, P. and {Knobel}, C. and {Kova{\v{c}}}, K. and {Le Borgne}, J. -F. and {Le Brun}, V. and {Maier}, C. and {Pell{\'o}}, R. and {Peng}, Y. and {Perez-Montero}, E. and {Presotto}, V. and {Silverman}, J.~D. and {Tanaka}, M. and {Tasca}, L.~A.~M. and {Tresse}, L. and {Vergani}, D. and {Almaini}, O. and {Barnes}, L. and {Bordoloi}, R. and {Bradshaw}, E. and {Cappi}, A. and {Chuter}, R. and {Cirasuolo}, M. and {Coppa}, G. and {Diener}, C. and {Foucaud}, S. and {Hartley}, W. and {Kamionkowski}, M. and {Koekemoer}, A.~M. and {L{\'o}pez-Sanjuan}, C. and {McCracken}, H.~J. and {Nair}, P. and {Oesch}, P. and {Stanford}, A. and {Welikala}, N.},
        title = "{Improved constraints on the expansion rate of the Universe up to z \raisebox{-0.5ex}\textasciitilde 1.1 from the spectroscopic evolution of cosmic chronometers}",
      journal = {jcap},
         year = 2012,
        month = aug,
       volume = {2012},
       number = {8},
          eid = {006},
        pages = {006},
          doi = {10.1088/1475-7516/2012/08/006},
archivePrefix = {arXiv},
       eprint = {1201.3609},
 primaryClass = {astro-ph.CO},
       adsurl = {https://ui.adsabs.harvard.edu/abs/2012JCAP...08..006M}
}

@ARTICLE{2010JCAP...02..008S,
       author = {{Stern}, Daniel and {Jimenez}, Raul and {Verde}, Licia and {Kamionkowski}, Marc and {Stanford}, S. Adam},
        title = "{Cosmic chronometers: constraining the equation of state of dark energy. I: H(z) measurements}",
      journal = {jcap},
         year = 2010,
        month = feb,
       volume = {2010},
       number = {2},
          eid = {008},
        pages = {008},
          doi = {10.1088/1475-7516/2010/02/008},
archivePrefix = {arXiv},
       eprint = {0907.3149},
 primaryClass = {astro-ph.CO},
       adsurl = {https://ui.adsabs.harvard.edu/abs/2010JCAP...02..008S}
}

@ARTICLE{2014RAA....14.1221Z,
       author = {{Zhang}, Cong and {Zhang}, Han and {Yuan}, Shuo and {Liu}, Siqi and {Zhang}, Tong-Jie and {Sun}, Yan-Chun},
        title = "{Four new observational H(z) data from luminous red galaxies in the Sloan Digital Sky Survey data release seven}",
      journal = {Research in Astronomy and Astrophysics},
         year = 2014,
        month = oct,
       volume = {14},
       number = {10},
          eid = {1221-1233},
        pages = {1221-1233},
          doi = {10.1088/1674-4527/14/10/002},
archivePrefix = {arXiv},
       eprint = {1207.4541},
 primaryClass = {astro-ph.CO},
       adsurl = {https://ui.adsabs.harvard.edu/abs/2014RAA....14.1221Z}
}

@article{Li:2019nux,
    author = "Li, En-Kun and Du, Minghui and Zhou, Zhi-Huan and Zhang, Hongchao and Xu, Lixin",
    title = "{Testing the effect of $H_0$ on $f\sigma_8$ tension using a Gaussian process method}",
    eprint = "1911.12076",
    archivePrefix = "arXiv",
    primaryClass = "astro-ph.CO",
    doi = "10.1093/mnras/staa3894",
    journal = "Mon. Not. Roy. Astron. Soc.",
    volume = "501",
    number = "3",
    pages = "4452--4463",
    year = "2021"
}

@article{Camarena:2023rsd,
    author = "Camarena, David and Marra, Valerio",
    title = "{The tension in the absolute magnitude of Type Ia supernovae}",
    eprint = "2307.02434",
    archivePrefix = "arXiv",
    primaryClass = "astro-ph.CO",
    month = "7",
    year = "2023"
}

@article{Chander:2025bml,
    author = "Chander, Lokesh and Singh, C. P.",
    title = "{Decaying vacuum energy, matter creation and cosmic acceleration}",
    eprint = "2504.09523",
    archivePrefix = "arXiv",
    primaryClass = "astro-ph.CO",
    month = "4",
    year = "2025"
}

@article{Riess:2019cxk,
    author = "Riess, Adam G. and Casertano, Stefano and Yuan, Wenlong and Macri, Lucas M. and Scolnic, Dan",
    title = "{Large Magellanic Cloud Cepheid Standards Provide a 1{\%} Foundation for the Determination of the Hubble Constant and Stronger Evidence for Physics beyond $\Lambda$CDM}",
    eprint = "1903.07603",
    archivePrefix = "arXiv",
    primaryClass = "astro-ph.CO",
    doi = "10.3847/1538-4357/ab1422",
    journal = "Astrophys. J.",
    volume = "876",
    number = "1",
    pages = "85",
    year = "2019"
}

@article{Jimenez:2023flo,
    author = "Jimenez, Raul and Moresco, Michele and Verde, Licia and Wandelt, Benjamin D.",
    title = "{Cosmic chronometers with photometry: a new path to H(z)}",
    eprint = "2306.11425",
    archivePrefix = "arXiv",
    primaryClass = "astro-ph.CO",
    doi = "10.1088/1475-7516/2023/11/047",
    journal = "JCAP",
    volume = "11",
    pages = "047",
    year = "2023"
}

@article{Moresco:2022phi,
    author = "Moresco, Michele and others",
    title = "{Unveiling the Universe with emerging cosmological probes}",
    eprint = "2201.07241",
    archivePrefix = "arXiv",
    primaryClass = "astro-ph.CO",
    doi = "10.1007/s41114-022-00040-z",
    journal = "Living Rev. Rel.",
    volume = "25",
    number = "1",
    pages = "6",
    year = "2022"
}

@article{Jimenez:2001gg,
    author = "Jimenez, Raul and Loeb, Abraham",
    title = "{Constraining cosmological parameters based on relative galaxy ages}",
    eprint = "astro-ph/0106145",
    archivePrefix = "arXiv",
    doi = "10.1086/340549",
    journal = "Astrophys. J.",
    volume = "573",
    pages = "37--42",
    year = "2002"
}

@article{PhysRevD.72.123519,
  title = {Comparison of the legacy and gold type Ia supernovae dataset constraints on dark energy models},
  author = {Nesseris, S. and Perivolaropoulos, L.},
  journal = {Phys. Rev. D},
  volume = {72},
  issue = {12},
  pages = {123519},
  numpages = {8},
  year = {2005},
  month = {Dec},
  publisher = {American Physical Society},
  doi = {10.1103/PhysRevD.72.123519},
  url = {https://link.aps.org/doi/10.1103/PhysRevD.72.123519}
}

@article{Gong:2007se,
    author = "Gong, Yan and Chen, Xuelei",
    title = "{Two Component Model of Dark Energy}",
    eprint = "0708.2977",
    archivePrefix = "arXiv",
    primaryClass = "astro-ph",
    doi = "10.1103/PhysRevD.76.123007",
    journal = "Phys. Rev. D",
    volume = "76",
    pages = "123007",
    year = "2007"
}

@article{Li:2024hrv,
    author = "Li, Jun-Xian and Wang, Shuang",
    title = "{A comprehensive numerical study on four categories of holographic dark energy models}",
    eprint = "2412.09064",
    archivePrefix = "arXiv",
    primaryClass = "astro-ph.CO",
    month = "12",
    year = "2024"
}

@ARTICLE{2001A&A...380....6G,
       author = {{Goliath}, M. and {Amanullah}, R. and {Astier}, P. and {Goobar}, A. and {Pain}, R.},
        title = "{Supernovae and the nature of the dark energy}",
      journal = {aap},
         year = 2001,
        month = dec,
       volume = {380},
        pages = {6-18},
          doi = {10.1051/0004-6361:20011398},
archivePrefix = {arXiv},
       eprint = {astro-ph/0104009},
 primaryClass = {astro-ph},
       adsurl = {https://ui.adsabs.harvard.edu/abs/2001A&A...380....6G}
}

@article{Scolnic:2021amr,
    author = "Scolnic, Dan and others",
    title = "{The Pantheon+ Analysis: The Full Data Set and Light-curve Release}",
    eprint = "2112.03863",
    archivePrefix = "arXiv",
    primaryClass = "astro-ph.CO",
    doi = "10.3847/1538-4357/ac8b7a",
    journal = "Astrophys. J.",
    volume = "938",
    number = "2",
    pages = "113",
    year = "2022"
}

@article{Rubin:2023jdq,
    author = "Rubin, David and others",
    title = "{Union Through UNITY: Cosmology with 2,000 SNe Using a Unified Bayesian Framework}",
    eprint = "2311.12098",
    archivePrefix = "arXiv",
    primaryClass = "astro-ph.CO",
    month = "11",
    year = "2023"
}

@article{DES:2024jxu,
    author = "Abbott, T. M. C. and others",
    collaboration = "DES",
    title = "{The Dark Energy Survey: Cosmology Results with {\ensuremath{\sim}}1500 New High-redshift Type Ia Supernovae Using the Full 5 yr Data Set}",
    eprint = "2401.02929",
    archivePrefix = "arXiv",
    primaryClass = "astro-ph.CO",
    reportNumber = "FERMILAB-PUB-23-0821-PPD, DES-2023-805",
    doi = "10.3847/2041-8213/ad6f9f",
    journal = "Astrophys. J. Lett.",
    volume = "973",
    number = "1",
    pages = "L14",
    year = "2024"
}

@ARTICLE{2009ApJ...700..331H,
       author = {{Hicken}, Malcolm and {Challis}, Peter and {Jha}, Saurabh and {Kirshner}, Robert P. and {Matheson}, Tom and {Modjaz}, Maryam and {Rest}, Armin and {Wood-Vasey}, W. Michael and {Bakos}, Gaspar and {Barton}, Elizabeth J. and {Berlind}, Perry and {Bragg}, Ann and {Brice{\~n}o}, Cesar and {Brown}, Warren R. and {Caldwell}, Nelson and {Calkins}, Mike and {Cho}, Richard and {Ciupik}, Larry and {Contreras}, Maria and {Dendy}, Kristi-Concannon and {Dosaj}, Anil and {Durham}, Nick and {Eriksen}, Kris and {Esquerdo}, Gil and {Everett}, Mark and {Falco}, Emilio and {Fernandez}, Jose and {Gaba}, Alejandro and {Garnavich}, Peter and {Graves}, Genevieve and {Green}, Paul and {Groner}, Ted and {Hergenrother}, Carl and {Holman}, Matthew J. and {Hradecky}, Vit and {Huchra}, John and {Hutchison}, Bob and {Jerius}, Diab and {Jordan}, Andres and {Kilgard}, Roy and {Krauss}, Miriam and {Luhman}, Kevin and {Macri}, Lucas and {Marrone}, Daniel and {McDowell}, Jonathan and {McIntosh}, Daniel and {McNamara}, Brian and {Megeath}, Tom and {Mochejska}, Barbara and {Munoz}, Diego and {Muzerolle}, James and {Naranjo}, Orlando and {Narayan}, Gautham and {Pahre}, Michael and {Peters}, Wayne and {Peterson}, Dawn and {Rines}, Ken and {Ripman}, Ben and {Roussanova}, Anna and {Schild}, Rudolph and {Sicilia-Aguilar}, Aurora and {Sokoloski}, Jennifer and {Smalley}, Kyle and {Smith}, Andy and {Spahr}, Tim and {Stanek}, K.~Z. and {Barmby}, Pauline and {Blondin}, St{\'e}phane and {Stubbs}, Christopher W. and {Szentgyorgyi}, Andrew and {Torres}, Manuel A.~P. and {Vaz}, Amili and {Vikhlinin}, Alexey and {Wang}, Zhong and {Westover}, Mike and {Woods}, Deborah and {Zhao}, Ping},
        title = "{CfA3: 185 Type Ia Supernova Light Curves from the CfA}",
      journal = {apj},
         year = 2009,
        month = jul,
       volume = {700},
       number = {1},
        pages = {331-357},
          doi = {10.1088/0004-637X/700/1/331},
archivePrefix = {arXiv},
       eprint = {0901.4787},
 primaryClass = {astro-ph.CO},
       adsurl = {https://ui.adsabs.harvard.edu/abs/2009ApJ...700..331H}
}

@ARTICLE{2012ApJS..200...12H,
       author = {{Hicken}, Malcolm and {Challis}, Peter and {Kirshner}, Robert P. and {Rest}, Armin and {Cramer}, Claire E. and {Wood-Vasey}, W. Michael and {Bakos}, Gaspar and {Berlind}, Perry and {Brown}, Warren R. and {Caldwell}, Nelson and {Calkins}, Mike and {Currie}, Thayne and {de Kleer}, Kathy and {Esquerdo}, Gil and {Everett}, Mark and {Falco}, Emilio and {Fernandez}, Jose and {Friedman}, Andrew S. and {Groner}, Ted and {Hartman}, Joel and {Holman}, Matthew J. and {Hutchins}, Robert and {Keys}, Sonia and {Kipping}, David and {Latham}, Dave and {Marion}, George H. and {Narayan}, Gautham and {Pahre}, Michael and {Pal}, Andras and {Peters}, Wayne and {Perumpilly}, Gopakumar and {Ripman}, Ben and {Sipocz}, Brigitta and {Szentgyorgyi}, Andrew and {Tang}, Sumin and {Torres}, Manuel A.~P. and {Vaz}, Amali and {Wolk}, Scott and {Zezas}, Andreas},
        title = "{CfA4: Light Curves for 94 Type Ia Supernovae}",
      journal = {apjs},
         year = 2012,
        month = jun,
       volume = {200},
       number = {2},
          eid = {12},
        pages = {12},
          doi = {10.1088/0067-0049/200/2/12},
archivePrefix = {arXiv},
       eprint = {1205.4493},
 primaryClass = {astro-ph.CO},
       adsurl = {https://ui.adsabs.harvard.edu/abs/2012ApJS..200...12H}
}

@article{Krisciunas:2017yoe,
    author = "Krisciunas, Kevin and others",
    title = "{The Carnegie Supernova Project I: Third Photometry Data Release of Low-Redshift Type Ia Supernovae and Other White Dwarf Explosions}",
    eprint = "1709.05146",
    archivePrefix = "arXiv",
    primaryClass = "astro-ph.IM",
    doi = "10.3847/1538-3881/aa8df0",
    journal = "Astron. J.",
    volume = "154",
    number = "5",
    pages = "211",
    year = "2017"
}

@article{Foley:2017zdq,
    author = "Foley, Ryan J. and others",
    title = "{The Foundation Supernova Survey: Motivation, Design, Implementation, and First Data Release}",
    eprint = "1711.02474",
    archivePrefix = "arXiv",
    primaryClass = "astro-ph.HE",
    doi = "10.1093/mnras/stx3136",
    journal = "Mon. Not. Roy. Astron. Soc.",
    volume = "475",
    number = "1",
    pages = "193--219",
    year = "2018"
}

@article{Brout:2022vxf,
    author = "Brout, Dillon and others",
    title = "{The Pantheon+ Analysis: Cosmological Constraints}",
    eprint = "2202.04077",
    archivePrefix = "arXiv",
    primaryClass = "astro-ph.CO",
    doi = "10.3847/1538-4357/ac8e04",
    journal = "Astrophys. J.",
    volume = "938",
    number = "2",
    pages = "110",
    year = "2022"
}

@article{Singirikonda:2020ieg,
    author = "Singirikonda, Haveesh and Desai, Shantanu",
    title = "{Model comparison of $\Lambda $CDM vs $R_h=ct$ using cosmic chronometers}",
    eprint = "2003.00494",
    archivePrefix = "arXiv",
    primaryClass = "astro-ph.CO",
    doi = "10.1140/epjc/s10052-020-8289-8",
    journal = "Eur. Phys. J. C",
    volume = "80",
    number = "8",
    pages = "694",
    year = "2020"
}

@article{Bouali:2019whr,
    author = "Bouali, Amine and Albarran, Imanol and Bouhmadi-L{\'o}pez, Mariam and Ouali, Taoufik",
    title = "{Cosmological constraints of phantom dark energy models}",
    eprint = "1905.07304",
    archivePrefix = "arXiv",
    primaryClass = "astro-ph.CO",
    doi = "10.1016/j.dark.2019.100391",
    journal = "Phys. Dark Univ.",
    volume = "26",
    pages = "100391",
    year = "2019"
}

\end{document}